\documentclass{aa}

\usepackage{graphicx}
\usepackage{epsfig}
\usepackage{txfonts}

\usepackage{units}
\usepackage{varioref}
\usepackage[breaklinks]{hyperref}

\usepackage{natbib}
\bibpunct{(}{)}{;}{a}{}{,} 

\labelformat{equation}{\textup{(#1)}}

\renewcommand{\figurename}{Fig.}

\newcommand{\dd}{\ensuremath{\mathrm d}}

\newcommand{\im}{\ensuremath{\mathrm i}}

\newcommand{\atanh}{\ensuremath{\mathrm{atanh}}}

\newcommand{\fvec}[1]{\mbox{\boldmath $#1$}}
\newcommand{\transpose}[1]{\ensuremath{#1^{\mathrm T}}}

\newcommand{\eg}{e.g.\ }

\newcommand{\SH}{SH2009}
\newcommand{\WS}{WS2013}

\newcommand{\hinv}{\ensuremath{h^{-1}}}
\newcommand{\Msun}{\ensuremath{M_\odot}}
\newcommand{\Mcrit}{\ensuremath{M^\mathrm{crit}_{200}}}

\newcommand{\rnu}{\ensuremath{r_{n\mathrm{u}}}}
\newcommand{\rnl}{\ensuremath{r_{n\mathrm{l}}}}
\newcommand{\rnul}{\ensuremath{r_{n\mathrm{u,l}}}}

\begin{document} 

   \title{Constrained correlation functions from the Millennium Simulation}

   \author{P. Wilking \and R. R\"oseler \and P. Schneider}

   \institute{Argelander-Institut f\"ur Astronomie, Universit\"at Bonn, Auf dem H\"ugel 71, 53121 Bonn, Germany\\
   \email{[pwilking,peter]@astro.uni-bonn.de}}

   \date{Received 16 February 2015 / Accepted 13 July 2015}

   \abstract
   {In previous work, we developed a quasi-Gaussian approximation for the likelihood of correlation functions that  incorporates fundamental mathematical constraints on correlation functions, in contrast to the usual Gaussian approach.
   The analytical computation of these constraints is only feasible in the case of correlation functions of one-dimensional random fields.}
   {In this work, we aim to obtain corresponding constraints in the case of higher dimensional random fields and test them in a more realistic context.}
   {We develop numerical methods of computing the constraints on correlation functions that are also applicable for two- and three-dimensional fields. To test the accuracy of the numerically obtained constraints, we compare them to the analytical results for the one-dimensional case. Finally, we compute correlation functions from the halo catalog of the Millennium Simulation, check whether they obey the constraints, and examine the performance of the transformation used in the construction of the quasi-Gaussian likelihood.}
   {We find that our numerical methods of computing the constraints are robust and that the correlation functions measured from the Millennium Simulation obey them. Even though the measured correlation functions lie well inside the allowed region of parameter space, i.e., far away from the boundaries of the allowed volume defined by the constraints, we find strong indications that the quasi-Gaussian likelihood yields a substantially more accurate description than the Gaussian one.}
   {}


\maketitle


\section{Introduction}
\label{sec:intro}

The two-point correlation function $\xi$ is been a very common tool in cosmology, although an increasing amount of astronomical literature deals with higher order statistics.
Whenever correlation function measurements are used in a Bayesian framework  to determine cosmological parameters, the probability distribution function (PDF) of the correlation function is needed. Usually, this likelihood, $\mathcal{L}(\xi)$, is assumed to be a multivariate Gaussian distribution (see, for example, an analysis of the correlation function of the cosmic microwave background by \cite{bib_COBE_likelihood_corr_function}, or common methods of baryon acoustic oscillations detection (e.g., by \citealt{bib_bao_labatie})).

However, the Gaussian approximation of $\mathcal{L}(\xi)$ is not necessarily well justified in all cases and may not always provide the level of precision required from statistical tools that are used to analyze state-of-the-art astronomical data, for example, non-Gaussianities in the cosmic shear likelihood were detected by \cite{bib_non_gaussianity_shear_likelihood}.
In the case of third-order cosmic shear statistics, however, \cite{bib_patrick_shear_likelihood} recently have found, at least in current state-of-the-art surveys, that a Gaussian likelihood is a reasonably good approximation. 
This agrees with results for the bispectrum covariance put forward by \cite{bib_sandra_bispectrum_covariance}.
As an additional remark, objections against the use of Gaussian likelihoods as a `safe default' have been raised in cases where one lacks knowledge of the exact form of the likelihood, as pointed out, for example, in power spectrum analyses by \cite{bib_carron_2012} and \cite{bib_sun_ps_likelihood}	.

A very strong argument against the Gaussianity of $\mathcal{L}(\xi)$ is the existence of fundamental constraints that stem from the non-negativity of the power spectrum and was put forward by \cite{bib_peter_jan_paper}, hereafter \SH{}. 
That correlation functions cannot take arbitrary values immediately implies that the Gaussian approximation cannot be fully correct, since a Gaussian distribution has infinite support.
To remedy this, one might be tempted to use a Gaussian likelihood for $\xi$ and include the constraints by simply incorporating priors that are zero outside the allowed region. However, as shown in previous work (see Figs.~4 and 5 in \SH{}), the shape of the distributions of $\xi$ are strongly affected by the constraints, even well inside the admissible range and thus a more comprehensive solution is needed.

Of course, it would be preferable to obtain the true PDF of $\xi$ analytically, which is feasible only for the uni- and bivariate cases, even assuming one-dimensional Gaussian random fields, as shown by \cite{bib_david_paper}.
Their results are a crucial ingredient of the quasi-Gaussian approach introduced in \cite{bib_my_paper}, hereafter \WS{}. There, we use the aforementioned constraints to transform the correlation function into an unconstrained variable, where the Gaussian approximation is expected to hold to higher accuracy. 
Using numerical simulations, we show that for the correlation functions of one-dimensional Gaussian fields, this `quasi-Gaussian transformation' performs very well, meaning that it transforms $\xi$ into a variable that is highly Gaussian. When we make  use of the analytical univariate $p(\xi)$ from \cite{bib_david_paper}, this transformation can then be exploited to construct the quasi-Gaussian likelihood for $\xi$. As presented in \WS{}, the new description of $\mathcal{L}(\xi)$ agrees well with the distributions obtained from simulations and has an impact on the results of Bayesian parameter estimation, as shown in their toy-model analysis.

To date, a major caveat of the quasi-Gaussian approach stems from the fact that the analytical computation of the constraints presented in \SH{} is only optimal for one-dimensional random fields. This severely limits the set of possible applications of the results presented in \WS{}. In \autoref{sec:constraints}, we develop numerical methods to compute the constraints on correlation functions that are also applicable to higher dimensional random fields, check their robustness, and compare the numerically obtained constraints to the analytical results for the one-dimensional case.
In \autoref{sec:mill}, we then apply the derived methods in an astrophysical context, i.e., to correlation functions measured from the halo catalog of the Millennium Simulation. 
We discuss some practical aspects of measuring $\xi$ and show that the correlation functions obtained from the simulation clearly obey the constraints. Furthermore, we examine the performance of the quasi-Gaussian transformation: By comparing the skewness and kurtosis of the transformed and the untransformed correlation functions, we argue that the quasi-Gaussian PDF is a better description of the likelihood of correlation functions than the Gaussian one.
We conclude with a brief summary and outlook in \autoref{sec:conclusions}.


\section{Numerical computation of the constraints on correlation functions}
\label{sec:constraints}

We consider the two-point correlation function of a random field $g(\vec x)$, which is defined as $\xi(\vec x, \vec y)=\left<g(\vec x)\ g^*(\vec y)\right>$. It is related to the power spectrum via Fourier transform. If assuming isotropy, this can be written as
\begin{equation}
 \xi(s) = \int\frac{\dd^n k}{(2\pi)^n}P(|\vec{k}|)\exp(\im\vec{k}\cdot\vec{x})
 = \int\frac{\dd k}{2\pi} k^{n-1}P(k)\ Z_n(ks),
 \label{eq:xi_FT}
\end{equation}
where $s\equiv |\vec x|$, and the dimensionality $n$ of the underlying random field determines the function $Z_n(\eta)$.
For a one-dimensional field, \autoref{eq:xi_FT} becomes a cosine transform; in the 2D case, $Z_2(\eta)=J_0(\eta)$ is the Bessel function of the first kind of zero order; and for a 3D random field, $Z_3(\eta)=j_0(\eta)$ is the spherical Bessel function of zero order.

As \SH{} show, correlation functions obey fundamental constraints, which arise from the non-negativity of the power spectrum and are best expressed in terms of the correlation coefficients $r_n\equiv \xi(s_n)/\xi(0)$.
As it turns out, the constraints can then be written in the form
\begin{equation}
 \rnl(r_1,r_2,\ldots,r_{n-1})\leq r_n\leq \rnu(r_1,r_2,\ldots,r_{n-1}), 
\end{equation}
meaning that the upper and lower boundaries on $r_n$ are functions of the $r_i$ with $i<n$.

\SH\ use the fact that the covariance matrix $C_{ij}=\langle g_i g_j^* \rangle =\xi_{|i-j|}$
(where $g_i=g(i\ \Delta x)$ for a one-dimensional random field evaluated at discrete grid points) has to be positive semi-definite,{} to explicitly calculate the constraints in the case of homogeneous, isotropic random fields, and show that the constraints they obtain are optimal for a one-dimensional random field, meaning that no stricter bounds exist for a general power spectrum. For higher dimensional fields, the bounds found for the one-dimensional case are still obeyed; however, owing to the isotropy of the field and the multidimensional integration in \autoref{eq:xi_FT}, tighter constraints hold that have to be computed numerically.

The procedure to obtain the optimal constraints numerically is outlined in \SH{}.  Rewriting \autoref{eq:xi_FT} and applying a quadrature formula for the integral yields
\begin{equation}
 r(s)\equiv \xi(s)/\xi(0) = \sum_{j=1}^K V_j\ Z_n(k_j s),
 \label{eq:r_Vj}
\end{equation}
where the coefficients fulfill $0\leq V_j\leq 1$ and $\sum V_j=1$. We note that this approximation becomes arbitrarily accurate as $K\rightarrow\infty$.

When measuring correlation coefficients for $N$ different separations $s_i$, each point $\fvec r=\left(r_1,r_2,\dots,r_N\right)$, with $r_i=\xi(s_i)/\xi(0)$, in this $N$-dimensional space can be written as a weighted sum along the curve $\fvec c(\lambda)=(Z_n(\lambda s_1), \dots,Z_n(\lambda s_N))$, where we used a continuous variable $\lambda$ with $0\leq \lambda < \infty$ instead of discrete wave numbers $k_j$:
\begin{equation}
 \fvec r=\sum_{j=1}^K V_j\ \fvec c(\lambda_j).
\end{equation}
Since $0\leq V_j\leq 1$ and $\sum V_j=1$, each point $\fvec r$ has to lie within the convex envelope of the curve $\fvec c(\lambda)$, which corresponds to the constraints on the correlation coefficients; for example, by constructing the convex envelope of the curve $\fvec c(\lambda)$ for two lags $(r_1, r_2)$ in the one-dimensional case, this reproduces the analytically known bounds $r_{2\mathrm{u,l}}(r_1)$.

As a result, to find the constraints only requires describing the convex envelope of the curve $\fvec c(\lambda)$. Unfortunately, there does not seem to be a general analytical solution for this problem, which means resorting  to numerical methods; for example, the qhull algorithm (\citealt{bib_qhull},  available at \url{http://www.qhull.org}) provides an efficient implementation for computing, among other things, convex hulls. It is, however, limited to inputs of dimensionality lower than nine, meaning that it is only applicable for a maximum number of separations of $N=8$. Although this is not a requirement for  computing  the constraints, we use equidistant lags throughout this work, denoting $s_n=n\ \Delta s$.

\begin{figure}
   \centering
   \resizebox{0.94\hsize}{!}{\includegraphics[keepaspectratio, angle=-90]{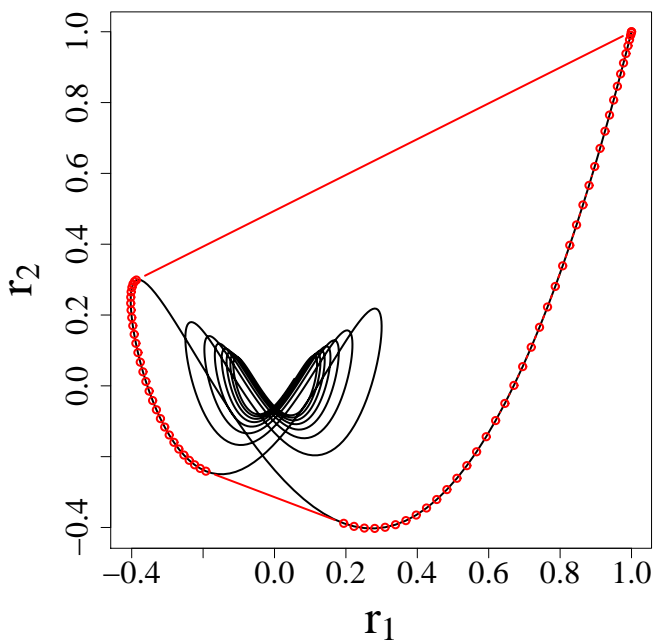}}
   \caption{Example of the curve $\fvec c(\lambda)$ for a two-dimensional random field in the $r_1-r_2$-plane, where $\fvec c(\lambda)= (J_0(\lambda),J_0(2\lambda))$. The red circles and the line connecting them show the convex hulls determined by qhull.}
   \label{fig:c_of_lambda}
\end{figure}

As an example of  determining  the constraints, \autoref{fig:c_of_lambda} shows the curve $\fvec c(\lambda)$ in the $r_1-r_2$-plane, plotted in black up to as high as $\lambda=50$ for illustrative purposes, as well as its convex hull.
For a given $r_1$, the upper and lower bounds on $r_2$ are given as intersections with the red hull. This method can, of course, be generalized to higher dimensions, \eg the determination of $r_{5\mathrm{u,l}}$ from $r_1,\dots,r_4$, where the convex hull is a hypersurface in a five-dimensional space.
Following this procedure, we developed a code to compute the constraints for one-, two-, and three-dimensional fields that we  use in our analysis of correlation functions measured from the Millennium Simulation in \autoref{sec:mill}.

\subsection{Comparison of the constraints for one-dimensional fields}
\label{sec:constraints_comparison}

Below, we  test the numerical method to obtain the constraints. To do so, we compare the numerical results to the analytically computed bounds. Since an analytical calculation of the constraints is only possible for one-dimensional random fields and equidistant lags, we limit our comparison to this case. Throughout this work, we use a gridded approach and  denote $\xi_n\equiv \xi(s_n)=\xi(n\ \Delta s)$, where $\Delta s=L/N$ is the separation between adjacent grid points, and $L$ denotes the length of the field.

There are several ways of testing our methods of computing the constraints: Most straightforward is to compare the constraints from the two methods directly, i.e., to compute the upper and lower bounds $\rnu$ and $\rnl$ both analytically and numerically, and check how much they differ.
An alternative approach involves the quasi-Gaussian transformation $r_n\rightarrow y_n$, which, as briefly explained in \autoref{sec:intro}, is a central ingredient of the quasi-Gaussian approximation for the likelihood of the correlation functions introduced in \WS{}:
\begin{equation}
 y_n=\atanh\frac{2r_n-r_{n\mathrm u}-r_{n\mathrm l}}{r_{n\mathrm u}-r_{n\mathrm l}}.
 \label{eq:r_to_y}
\end{equation}
Since this transformation is the main application for the constraints, it can -- and should be -- applied as a means to compare the analytically and numerically obtained bounds, namely by using the different sets of constraints in the transformation and comparing the resulting $y_n$.

As previously described, the constraints on $r_n$ are functions of the correlation coefficients with lower lags, and as such, we need input values for $r_1, \ldots, r_{n-1}$  to be able to compute and compare the different $\rnl$ and $\rnu$. Again, two possibilities arise:
To provide input values that are close to `real-life' applications, we can use realizations of correlation coefficients obtained from numerical simulations (see \WS{} for an efficient way to generate realizations of the correlation function of a one-dimensional Gaussian random field). However, this obviously requires assumptions about the underlying random field and, in particular, its power spectrum.
Consequently, a more general approach is to draw the input correlation coefficients for  computing the constraints randomly, i.e., from a uniform distribution over the allowed range, $r_n \in \mathopen{]}\rnl,\rnu\mathclose{[}$.
Due to the nature of the constraints, this is an iterative procedure, meaning that one has to draw $r_1\in \mathopen{]}r_{1\mathrm{l}},r_{1\mathrm{u}}\mathclose{[}$, compute $r_{2\mathrm{l,u}}$ from this $r_1$, then draw $r_2\in \mathopen{]}r_{2\mathrm{l}},r_{2\mathrm{u}}\mathclose{[}$  to determine $r_{3\mathrm{l,u}}$, and so on.

\begin{figure}
   \centering
   \resizebox{0.94\hsize}{!}{\includegraphics[keepaspectratio, angle=-90]{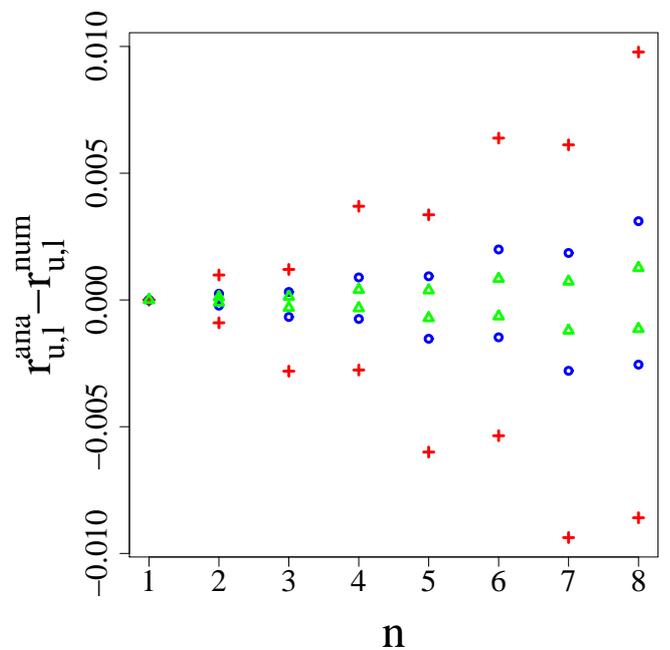}}\hfill
   \caption{Difference between the analytically and numerically obtained bounds, averaged over 500 realizations. The upper three sets of points correspond to the difference $r_\mathrm{u}^\mathrm{ana}-r_\mathrm{u}^\mathrm{num}$, whereas the ones with negative values show $r_\mathrm{l}^\mathrm{ana}-r_\mathrm{l}^\mathrm{num}$. Furthermore, the different symbols denote the number of steps used to sample the convex hull of the curve $\fvec c(\lambda)$ for values of $0\leq\lambda\leq 2\pi$, namely 100 (red crosses), 200 (blue circles), and 300 (green triangles) steps.}
   \label{fig:constraints_diff}
\end{figure}

A comparison of the analytically and numerically obtained bounds $\rnu$ and $\rnl$ is shown in \autoref{fig:constraints_diff}, where we plot the differences $r_\mathrm{u,l}^\mathrm{ana}-r_\mathrm{u,l}^\mathrm{num}$ as functions of $n$. For each bound $\rnul$, the required input values of the correlation coefficients $r_i$ with $i<n$ are drawn uniformly, as previously described.
To perform a statistically significant check, this procedure is repeated 500 times, meaning that we generate 500 realizations of the input correlation coefficients and compute the upper and lower bounds both numerically and analytically for each realization. The values plotted in the figure are obtained by averaging the difference between the analytical and the numerical values over the 500 realizations. In addition, we also investigate how much impact the sampling of the convex hull of the curve $\fvec c(\lambda)$ has on the accuracy of the numerical bounds.
It turns out that it is sufficient in all cases to sample the curve $\fvec c(\lambda)$ for values of $0\leq\lambda\leq 2\pi$,
since going to higher values of $\lambda$ has no impact on the volume within the convex hull because of the periodicity of the function $Z_n(ks)$ in \autoref{eq:xi_FT}.
By way of comparison, we vary the sampling rate of the convex hull; i.e., we sample it using 100 (red crosses), 200 (blue circles), and 300 (green triangles) steps.
The upper three sets of points show the difference $r_\mathrm{u}^\mathrm{ana}-r_\mathrm{u}^\mathrm{num}$ between the analytical and numerical upper bounds, whereas the lower ones depict the deviation between the lower bounds, i.e., $r_\mathrm{l}^\mathrm{ana}-r_\mathrm{l}^\mathrm{num}$.

Three conclusions can be drawn from \autoref{fig:constraints_diff}: First, the deviation of the numerically obtained bounds from the analytical ones shows a tendency to grow with $n$, which is to be expected, since the sampling of the convex hull becomes more challenging with increasing dimensionality. (The numerical and analytical bounds on $r_1$ do not differ at all, since $r_{1\mathrm{u,l}}=\pm 1$.)
Second, the impact of this sampling has a strong impact on the accuracy of the numerical calculation of the bounds; namely, the difference between the numerical and the analytical results decreases by about a factor of three when doubling the number of steps used for the convex hull sampling.
Actually, this sampling is the limiting factor for the accuracy of the numerical bounds, as can be seen from the third observation: In the case of the upper bounds, the numerical results are systematically smaller than the exact analytical values, whereas for the lower bounds, the numerical values are too high.
This effect is an expected consequence of the non-continuous approximation for the smooth hull; as a result of convexity, the hyperplanes that link the points used to sample the hull always have to be located inside the hull.

In summary, the accuracy of the numerical constraints can be increased by improving the sampling of the hull. While a larger number of steps would, presumably, improve the results even further, using more than 300 steps for the sampling becomes impractical because of the computational costs. However, as the following tests   demonstrate, using 300 steps seems sufficiently accurate.

As mentioned before, another important check for the accuracy of the numerical methods that are used to compute the bounds is to apply the quasi-Gaussian transformation $r_n\rightarrow y_n$ and to compare the resulting $y_n$. Below, we adopt correlation coefficients from simulations instead of uniformly drawn ones as input for the computation of the bounds. In particular, we use 500 realizations of the correlation function on a one-dimensional Gaussian field of length $L$ with $N=32$ grid points and a Gaussian power spectrum, where the field length and the width of the power spectrum are related by $L k_0=80$. (For details on the simulations used in this work, we refer to \WS{}.)

\begin{figure}
   \centering
   \resizebox{0.95\hsize}{!}{\includegraphics[keepaspectratio, angle=-90]{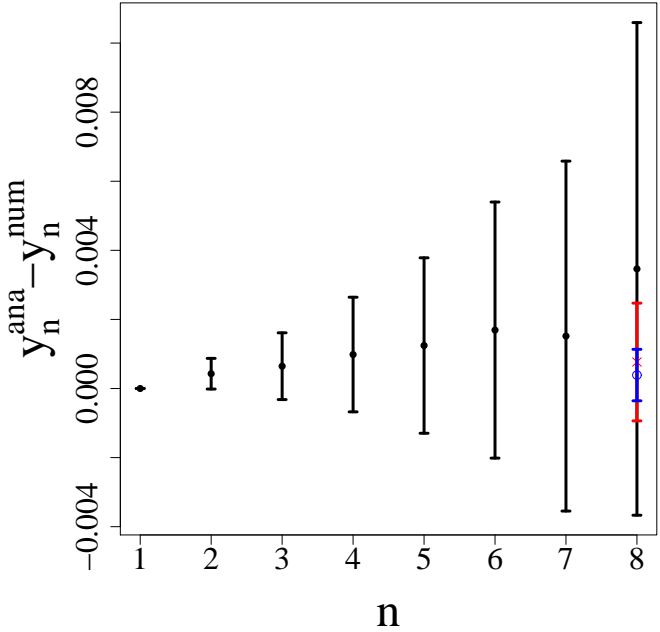}}\hfill
   \caption{Difference between the $y_n$ computed using the analytically and the numerically obtained bounds, averaged over 500 realizations, with error bars showing the standard deviations. The input values for the computation of the bounds used in the transformation $r_n\rightarrow y_n$ stem from 500 simulated realizations of the correlation function on a one-dimensional Gaussian field of length $L$ with $N=32$ grid points and a Gaussian power spectrum of width $k_0$, with $L k_0=80$. In the case of the black, solid points, the convex hull of the curve $\fvec c(\lambda)$ is sampled using 100 points for the range $0\leq\lambda\leq 2\pi$. By way of comparison, we show the corresponding results for sampling rates of 200 (red cross) and 300 (blue circle) in the case $n=8$.}
   \label{fig:y_diff}
\end{figure}

For each simulated realization of the correlation coefficients, we compute the bounds $\rnu$ and $\rnl$ for each $n$, both numerically and analytically, and use them to transform $r_n$ to $y_n$, as defined in \autoref{eq:r_to_y}. To compare the resulting values for $y_n$, we plot the differences $y_n^\mathrm{ana}-y_n^\mathrm{num}$ as a function of $n$ in \autoref{fig:y_diff}. Here, the plotted values are the average over the 500 realizations, while the error bars denote the standard deviations.
For the sake of clarity, we only show the impact of the number of steps used in the convex hull sampling for the case $n=8$, where the standard deviations are largest.

The accuracy of the numerical approximation again shows a strong dependence on the number of steps used to sample the convex hull. Nevertheless, the difference between the values of $y_n,$ computed using the analytical and the numerical bounds, becomes very small when using 300 steps. As a result,  we conclude that the problem of the non-continuous approximation of the curve $\fvec c(\lambda)$ and its convex hull can be tackled and that using 300 steps in the sampling yields sufficiently accurate bounds.


\section{Application to the Millennium Simulation}
\label{sec:mill}

So far, our studies about the constraints on correlation functions and the quasi-Gaussian likelihood have been performed in a general, mathematical framework. In this section, we  investigate our results in a more astrophysical context by applying them to cosmological correlation functions measured in the Millennium Simulation (\citealt{bib_mill_springel}).
Thus, we aim to check the relevance of the constraints that originally stem from purely mathematical properties, since they are based on the fact that $\xi$ is the Fourier transform of a positive quantity (the power spectrum), in a more practical context, where $\xi$ is measured using an estimator.
The size of the Millennium Simulation enables us to easily measure multiple realizations of $\xi$, thereby providing an approximate determination of the underlying probability distribution and, consequently, a statistical analysis.

\subsection{Computing correlation functions}
\label{sec:computing_xi}

Below, we compute the correlation function of dark matter halos in the Millennium Simulation.  Because we are not interested in redshift evolution, we only  use  the halo catalog from the $z=0$ simulation snapshot, from which we then select typical galaxy-mass halos by choosing a mass cut $\Mcrit > \unit[10^{12}\ \hinv]{\Msun}$, which yields a total number of $\sim 440\ 000$ halos.
However, to perform a statistical analysis, we require different realizations of the correlation function. For this reason, as a first attempt, we divide the full simulation cube into 1000 subcubes of volume $50^3\ \left(\unit[\hinv]{Mpc}\right)^3$ and measure $\fvec\xi$ in each of the subcubes.

Along with the halo catalog from the simulation we also need a random catalog, so, for each subcube, we draw halo positions uniformly. We then determine the number of halo pairs for given pair separations in both the data and the random catalog, as well as the cross-correlation.
From the count rates $DD(s)$, $RR(s)$, $DR(s)$ (normalized to account for different numbers of halos in both the random catalog and the halo catalog from each subcube) at different pair separations $s$, we compute $\fvec\xi$ using an estimator. While Landy-Szalay (LS) is the most widely used   estimator (\citealt{bib_cf_landy_szalay}), we also aim to test the impact of the choice of estimator on the constraints, and adapt the following common set of estimators from \cite{bib_vm_cf_estimators}:

\begin{tabular}{lr}
 $\xi_\mathrm{PH}=\frac{DD}{RR}-1$, & \cite{bib_cf_peebles_hauser};\\[1ex]
 $\xi_\mathrm{Hew}=\frac{DD-DR}{RR}$, & \cite{bib_cf_hewett};\\[1ex]
 $\xi_\mathrm{DP}=\frac{DD}{DR}-1$, & \cite{bib_cf_davis_peebles};\\[1ex]
 $\xi_\mathrm{H}=\frac{DD\times RR}{DR^2}-1$, & \cite{bib_cf_hamilton};\\[1ex]
 $\xi_\mathrm{LS}=\frac{DD-2DR+RR}{RR}$, & \cite{bib_cf_landy_szalay}.
\end{tabular}

As mentioned in \autoref{sec:constraints}, to calculate and test the constraints, we measure the correlation function at equidistant lags, i.e., $\xi_n\equiv \xi(n\cdot\Delta s)$, where the maximum number of lags is $n=8$, an effect of the limitations of the numerical computation of the constraints.

The size of the random catalog also merits some discussion:\ Ideally, it should be infinitely large, i.e., $N_\mathrm{rand}\rightarrow\infty$. However, the computation of the pair separations is the most time-consuming step in the calculation of $\fvec\xi$ and, consequently, the number of halos in the random catalogs for each of the 1000 subcubes is subject to practical limitations. We study the impact of the random catalog size in \autoref{fig:xi_of_n_nrand}:
\begin{figure}
   \centering
   \resizebox{0.95\hsize}{!}{\includegraphics[keepaspectratio, angle=-90]{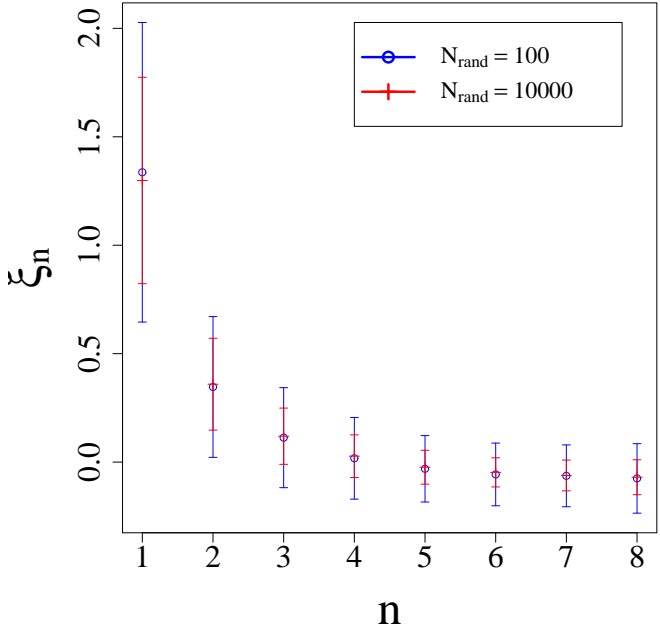}}\hfill
   \caption{Correlation function from 1000 subcubes of the Millennium Simulation, computed using the LS estimator. The points and error bars show the correlation function $\xi_n\equiv\xi(n\cdot\Delta s)$ for $\Delta s = \unit[5\ \hinv]{Mpc}$ (see text for details) averaged over the 1000 subcubes of side length $50\ \unit[\hinv]{Mpc}$, as well as the standard deviation. For the blue circles, the random catalog for each subcube contains 100 halos, as opposed to 10000 halos for the red crosses.}
   \label{fig:xi_of_n_nrand}
\end{figure}
Here, we show the correlation function for an exemplary choice of lags with $\Delta s=\unit[5\ \hinv]{Mpc}$, i.e., we measure $\xi_1,\xi_2,\ldots,\xi_8$ at lags of $5, 10, \ldots \unit[40\ \hinv]{Mpc}$. In practice, we need to allow for a range of pair separations to obtain sufficiently large numbers of pairs. To do this, we adapt a bin size of width $\unit[1\ \hinv]{Mpc}$, so, for example, in the computation of $\xi_1$ we use all pairs with separations ranging from 4.5 to $\unit[5.5\ \hinv]{Mpc}$.
For the auto-correlation $\xi_0$, i.e., the correlation function at zero lag (which we do not plot in the figure, but which is required for the calculation of the constraints), we count all pairs with very small separations, e.g., $s\leq\unit[1\ \hinv]{Mpc.}$ (We refer to the next section for a discussion on the measurement of $\xi_0$ and on the choice of lags.)
The points and error bars show the mean and standard deviation over the 1000 subcubes of side length $50\ \unit[\hinv]{Mpc}$; we use the LS estimator, which has been shown to be less sensitive to the size of the random catalog than others (see \citealt{bib_cf_estimators_comparison}).
For the blue circles, a small random catalog ($N_\mathrm{rand}=100$ halos for each subcube) was used, whereas the choice of $N_\mathrm{rand}=10000$ for the red crosses results in noticeably smaller standard deviations over the 1000 realizations, at the cost of a longer computation time.

Hence, we aim to find a trade-off between those two values. First, although the mean of the correlation functions for the two random catalog sizes plotted in \autoref{fig:xi_of_n_nrand} do not seem to differ very much at first sight, choosing the catalog size as small as $N_\mathrm{rand}=100$ is a quite extreme case. This is because a large fraction of the realizations yield a diverging auto-correlation $\xi_0$ as a result of the  count rates in $RR$ or $DR$ being zero, at least when measuring $\xi_0$ as previously described. However, when increasing the random catalog to 1000 halos per subcube, the mean correlation function for non-zero lag shows a deviation of only $\sim \unit[1]{\%}$ compared to the mean $\xi$ for $N_\mathrm{rand}=10000$ (and even here, about a tenth of the realizations show a diverging $\xi_0$).
At $N_\mathrm{rand}=5000$, no such problems occur, while even the error bars, as plotted in the figure, become indistinguishable from those at $N_\mathrm{rand}=10000$. This means that a random catalog size of 5000 is a reasonable trade-off between accuracy and computational expenses.

An additional observation from \autoref{fig:xi_of_n_nrand} is that $\xi$ becomes negative for larger lags, i.e., around $\unit[20-25\ \hinv]{Mpc}. $ The reason for this is an integral constraint (see, for example,  \citealt{bib_cf_landy_szalay}) that arises when measuring correlation functions in finite volumes, where the global mean density is unknown and is usually approximated by the mean observed density.

\begin{figure}
   \centering
   \resizebox{0.95\hsize}{!}{\includegraphics[keepaspectratio, angle=-90]{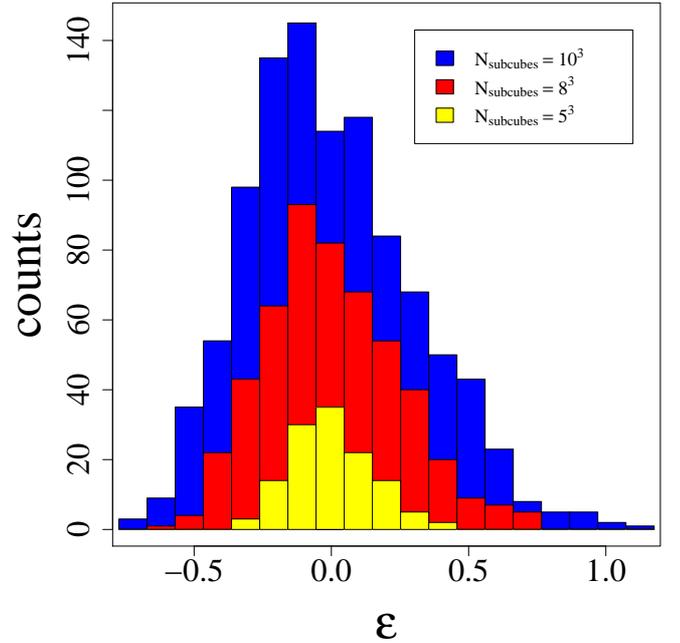}}\hfill
   \caption{Histograms of the distributions of overdensities in the subcubes of the simulation volume, where $\epsilon$ denotes the number overdensity of halos in the subcube relative to the mean halo number density in the whole simulation box, see \protect\autoref{eq:epsilon}. The colors indicate the number of subcubes the simulation box was divided into.}
   \label{fig:hists_subcubes}
\end{figure}

\begin{figure}
   \centering
   \resizebox{0.95\hsize}{!}{\includegraphics[keepaspectratio, angle=-90]{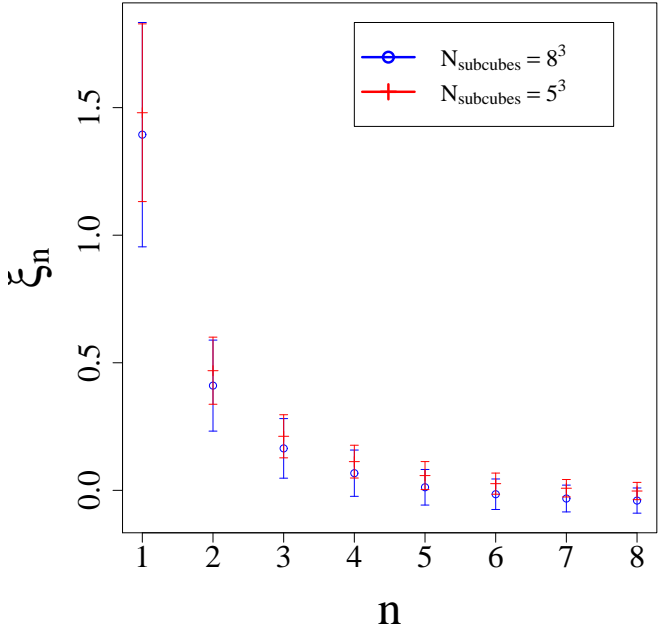}}
   \caption{Correlation function from the Millennium Simulation, computed using the LS estimator, a random catalog size of 5000, and a lag separation of $\Delta s=\unit[5\ \hinv]{Mpc}$. The points and error bars show the mean and standard deviation computed over the subcubes of the simulation, where the simulation box was sliced into $8^3$ subcubes for the blue data points, as opposed to $5^3$ for the red ones.}
   \label{fig:xi_of_n_subcubes}
\end{figure}

In our case, one way to assess this issue is to decrease the number of subcubes in our analysis, and make them larger and more representative for the whole box, while at the same time measuring $\xi$ at the same lags as before. 
Beside lessening the impact of the integral constraint on small lags, this has two additional effects: First, with fewer subcubes, we obtain fewer realizations of $\xi$, which can pose a challenge for a statistical analysis, and second, as the number of halo pairs per cube becomes larger, the scatter over the measured realizations of $\xi$ decreases.
To decide on the number of subcubes necessary to make the subcubes as representative as possible for the whole simulation volume, we estimate the overdensity $\epsilon$ in each subcube by comparing the mean number density in the subcube to the one from the whole simulation volume,
using\begin{equation}
 \bar{n}_\mathrm{sub} = \bar{n}_\mathrm{box} \left(1+\epsilon\right),
 \label{eq:epsilon}
\end{equation}
and examine the distributions of $\epsilon$ by plotting them as histograms in \autoref{fig:hists_subcubes}.
Here, we slice the simulation volume into different numbers of subcubes and compute the overdensity in each subcube. It is clear that the distribution $p(\epsilon)$ is very broad for the value $N_\mathrm{subcubes}=10^3$ used so far, and, as expected, it becomes quite narrow for the case of $5^3$ subcubes, which indicates that the integral constraint does not pose a large problem in this case. The resulting correlation functions for the cases of $8^3$ and $5^3$ subcubes (and otherwise the same parameters as before, i.e., same lags and random catalog size) are shown in \autoref{fig:xi_of_n_subcubes}.
As can be seen, slicing the simulation volume into $5^3$ subcubes yields reasonable results, i.e., a non-negative correlation functions with a sufficiently small variance.

Finally, we briefly evaluate the choice of estimator. To do so, instead of measuring $\xi_n$ at a few different lags $n$, it is advisable to compute $\xi(s)$ for all lags in each subcube. Since this is obviously not practicable, we compute it at a high number of lags, meaning that we divide the range of pair separations into adjacent bins of width $\unit[0.2\ \hinv]{Mpc}$.

The correlation functions (average over the $5^3$ subcubes and using a fixed size of random catalog for each subcube, namely $N_\mathrm{rand}=5000$) are shown in \autoref{fig:xi_of_s_est}.
Here, the different colored lines denote the five  estimators, and the gray-shaded region depicts the standard deviation over the 125 realizations in the case of the most commonly used LS estimator. For clarity, in the left panel, we plot scales from 8 to $\unit[40\ \hinv]{Mpc}$, whereas the right panel shows the correlation function for very small lags, i.e., $4-\unit[8\ \hinv]{Mpc}$. Clearly, the numerous estimators yield very similar results, in particular compared to the standard deviation of the 125 realizations.

\subsection{Testing the constraints}
\begin{figure*}
   \resizebox{0.44\hsize}{!}{\includegraphics[keepaspectratio, angle=-90]{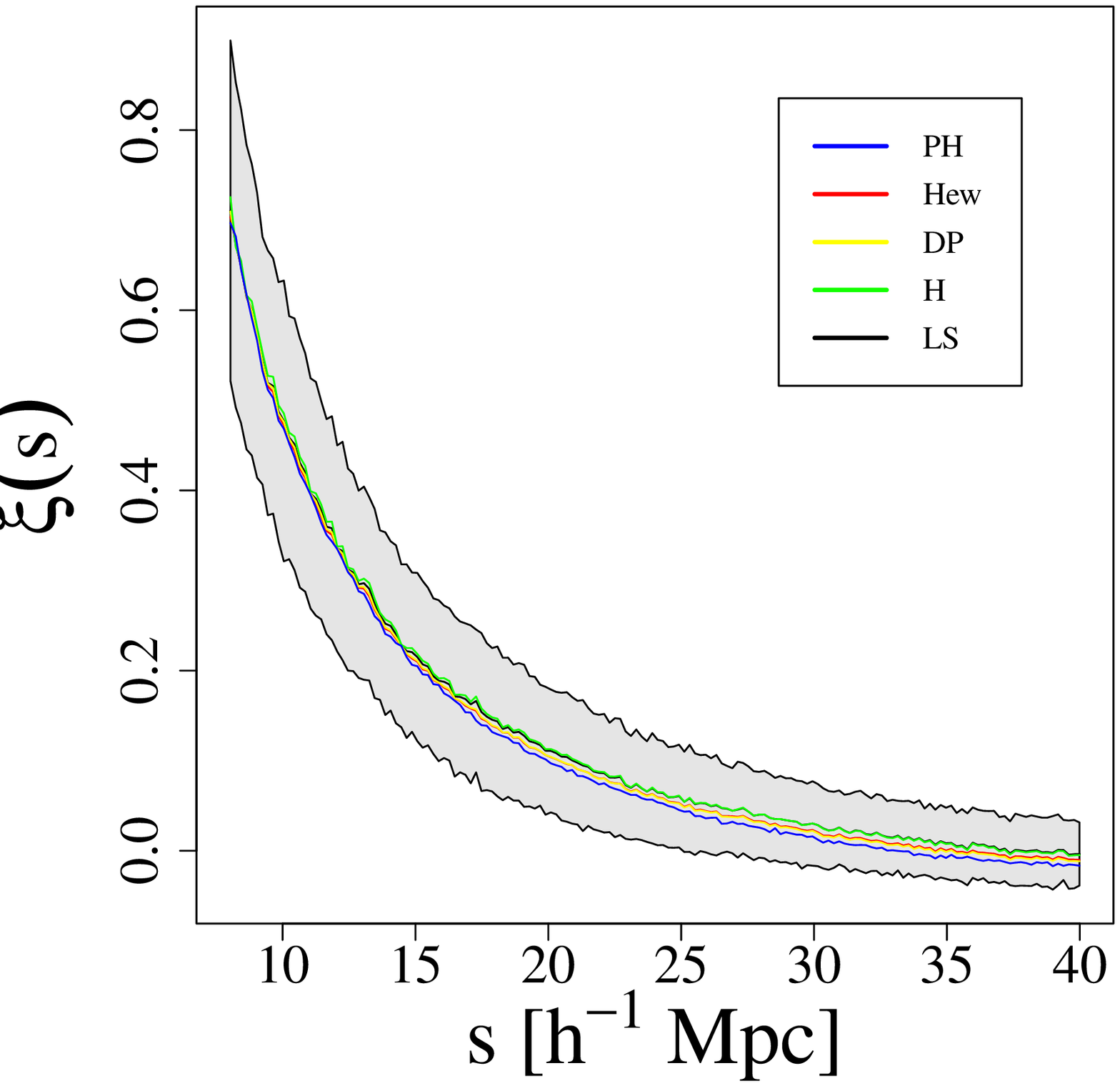}}\hfill
   \resizebox{0.44\hsize}{!}{\includegraphics[keepaspectratio, angle=-90]{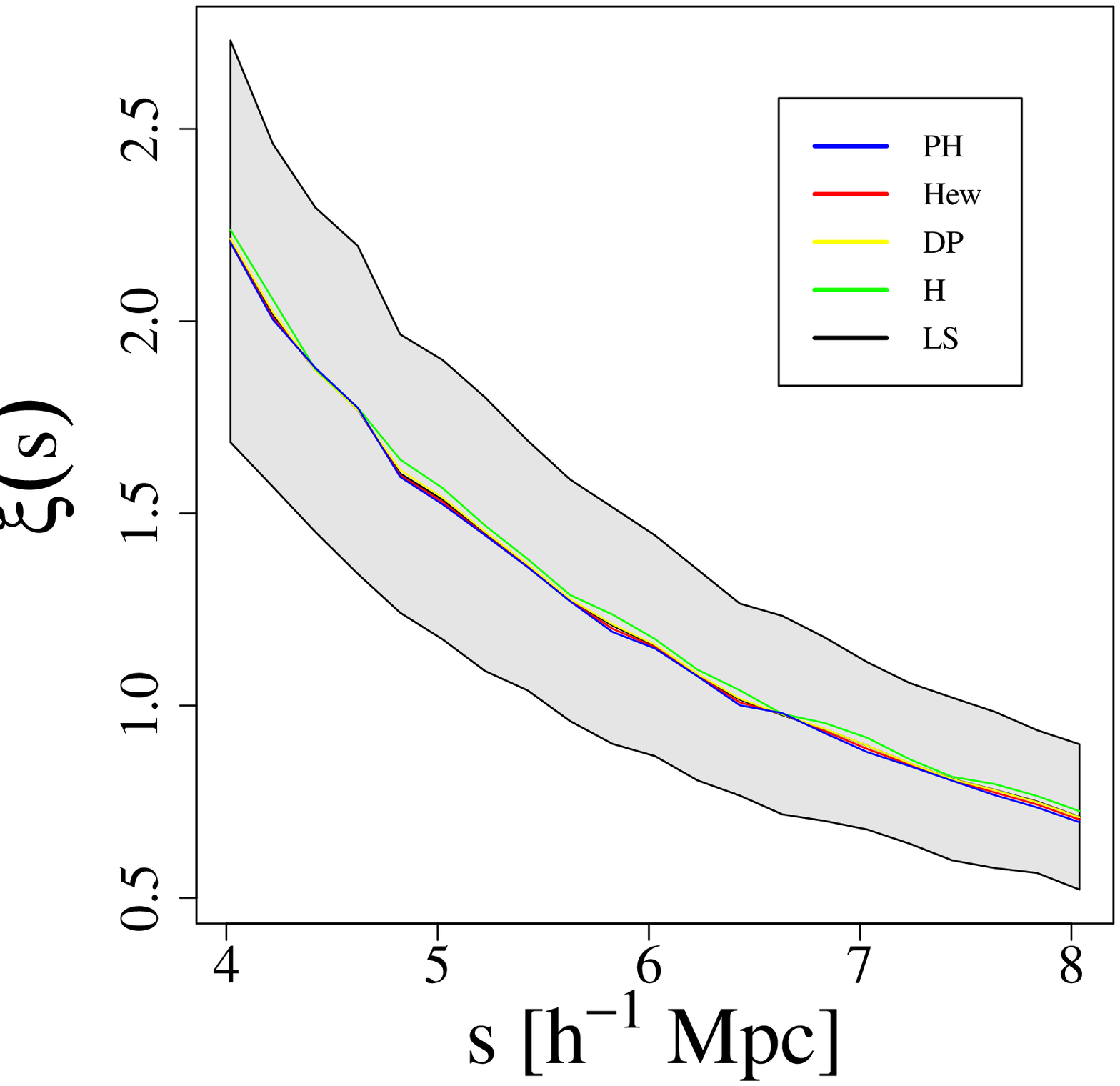}}
   \caption{Correlation function measured for `all' lags (see text for details) as a function of the pair separation $s$, with a random catalog size of $N_\mathrm{rand}=5000$. The lines denote the mean $\xi(s)$ measured in the 125 subcubes using the different estimators listed in \protect\autoref{sec:computing_xi}, and the gray shaded region shows the standard deviation. In the left panel, the $s$-range from 8 to $\unit[40\ \hinv]{Mpc}$ is plotted, and the right panel shows the results for very small lags, from 4 to $\unit[8\ \hinv]{Mpc}$.}
   \label{fig:xi_of_s_est}
\end{figure*}
In this section, we investigate whether the correlation functions computed from the halo catalog of the Millennium Simulation obey the numerically obtained constraints.
While we make the case in \autoref{sec:constraints_comparison}, that using 300 points to sample the hull yields sufficiently good agreement between the numerical and the exact analytical values of the constraints, we have to restrict ourselves to 270 points in the case of a 3D random field owing to the computational costs. Although the convex hull only has to be computed once and can then be used to determine the constraints for all sets of correlation coefficients, sampling the hull for a 3D random field with the given accuracy poses memory problems for the qhull algorithm, and  is beyond the scope of this research.
However, this does not pose a problem: When we compare the accuracy of the numerical constraints, as plotted in \autoref{fig:constraints_diff}, it is apparent that the improvement in accuracy, when going from 200 to 300 steps, is far smaller than the one from 100 to 200, and thus we expect  270 points to be accurate enough.

To test the constraints, for each realization we compute  the correlation coefficients $r_n\equiv \xi_n/\xi_0$ as well as the upper and lower bounds, $\rnu$ and $\rnl$.
It turns out that the width of the $\xi_0$-bin has a strong influence, in particular on the width of the distributions of the correlation functions $r_n$. For example, we first choose a relatively broad bin, i.e., we measure $\xi_0$ by averaging over all pair separations from 0 to $\unit[2\ \hinv]{Mpc. }$ This choice is primarily motivated by the fact that increasing the spread of the correlation coefficients over the 125 realizations allows us to test how close to the edges of the allowed region the $r_n$ move. Toward the end of this section, we  study the impact of the width of the $\xi_0$-bin in more detail.

\begin{figure*}
   \resizebox{0.44\hsize}{!}{\includegraphics[keepaspectratio, angle=-90]{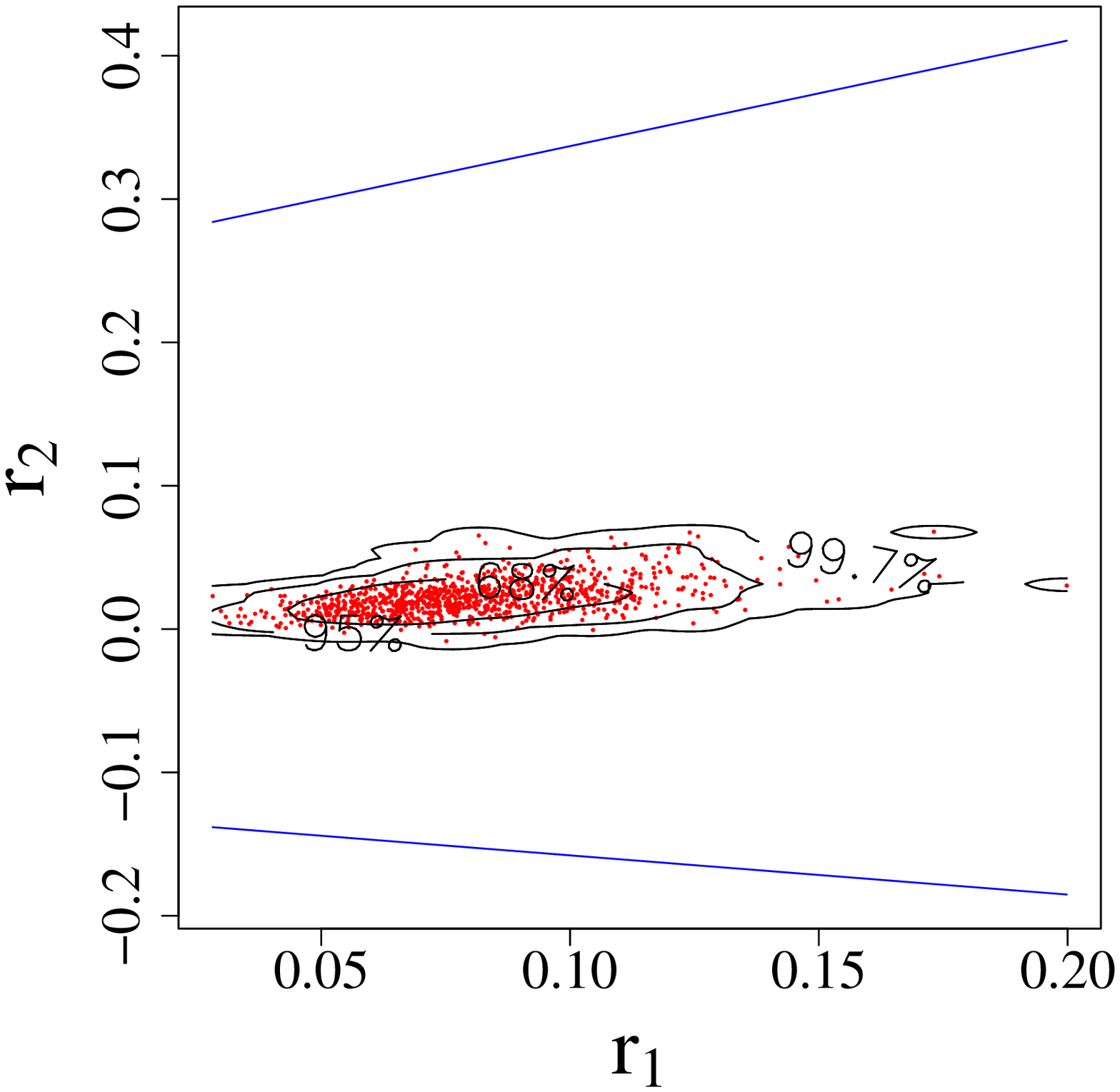}}\hfill
   \resizebox{0.44\hsize}{!}{\includegraphics[keepaspectratio, angle=-90]{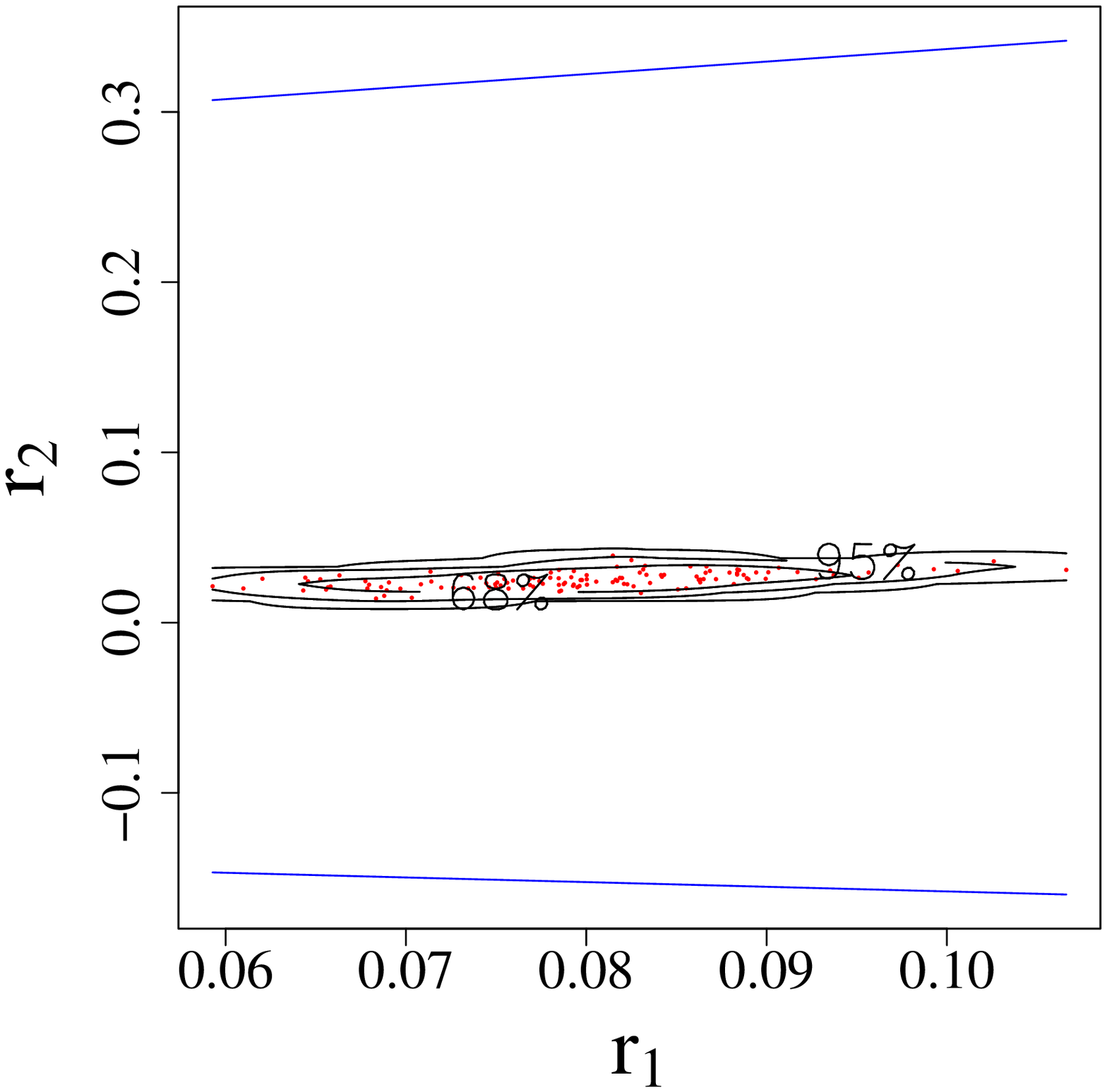}}
   \caption{Correlation coefficients $r_1$ and $r_2$ measured from the halo catalogs in the subcubes of the Millennium Simulation, using the LS estimator, where we slice the simulation volume into 1000 subcubes for the left panel and 125 for the right one, and the random catalog for each subcube contains 5000 halos. In both cases, we measure $\xi$ at lags of separation $\Delta s=\unit[5\ \hinv]{Mpc}$, and use all halo pairs with pair separations of 0 to $\unit[2\ \hinv]{Mpc}$ to compute $\xi_0$. The red dots show the 1000 (125) realizations, while the black lines are iso-density contours that contain the given percentages of the realizations.
   The upper and lower constraints $r_{2\mathrm{u,l}}(r_1)$, computed individually for each realization of $r_1$, are shown as blue lines.}
   \label{fig:contours_r1r2}
\end{figure*}

One question that arises is how to visualize the constraints. The simplest approach to this is to use scatter-plots, with dots for the individual realizations. An example in the $r_1-r_2$-plane is shown in \autoref{fig:contours_r1r2}.
Here, the red dots show the different realizations of $r_1$ and $r_2$, computed for the subcubes using the LS estimator. Additionally, we plot iso-density contours that contain 68, 95, and $\unit[99.7]{\%}$ of these realizations.
For the lefthand panel, we sliced the simulation volume into 1000 subcubes, as opposed to 125 for the righthand panel. As explained in the previous section, the higher number of subcubes greatly increases the spread of the correlation functions, which can also  be clearly observed in $r$-space. (Even for the high number of subcubes, the integral constraint is expected to be negligible for the correlation functions at small lags.)
In both panels, the upper and lower blue lines represent the constraints, i.e.,  $r_{2\mathrm{u,l}}(r_1)$, which we compute numerically for each realization of $r_1$ shown in the figure and plot as connected lines.
All realizations clearly lie well inside the constraints,  particularly when compared to results for purely Gaussian (one-dimensional) random fields with fiducial power spectra (see, for example, similar figures in related works, such as Fig.\ 3 from  \WS{}).

As an additional way of depicting the constraints, we apply a part of the quasi-Gaussian transformation from \autoref{eq:r_to_y}  to map the allowed range of the correlation coefficients to $(-1,+1)$, namely by transforming the correlation coefficients $r_n$ to 
\begin{equation}
 x_n=\frac{2r_n-r_{n\mathrm u}-r_{n\mathrm l}}{r_{n\mathrm u}-r_{n\mathrm l}}.
 \label{eq:r_to_x}
\end{equation}

To  visualize the constraints more clearly, we  use  a modified version of box-and-whisker plots, meaning that we display our samples $\lbrace r_n\rbrace$ and $\lbrace x_n\rbrace$ as boxes, where the upper and lower borders show the first and third quartiles of the sample, i.e., the values that split off the upper and lower $\unit[25]{\%}$ of the data. As is common practice, we also show the sample median (i.e., the second quartile) as a line inside the box, as well as two whiskers. Usually the ticks at the end of the whiskers denote the minimum and maximum
of the data (in the most widely used type of box-and-whisker plot). Here, we  use them to denote the upper and lower constraints: Since $r_{n\mathrm{u,l}}$ are functions of all $r_i$ with $i<n$, we show the mean $\rnu$ and $\rnl$ over all realizations for plots in $r$-space. For the transformed values $x_n$, the bounds are simply $\pm 1$, so there is no need to average over the realizations.

\begin{figure*}
   \resizebox{0.44\hsize}{!}{\includegraphics[keepaspectratio, angle=-90]{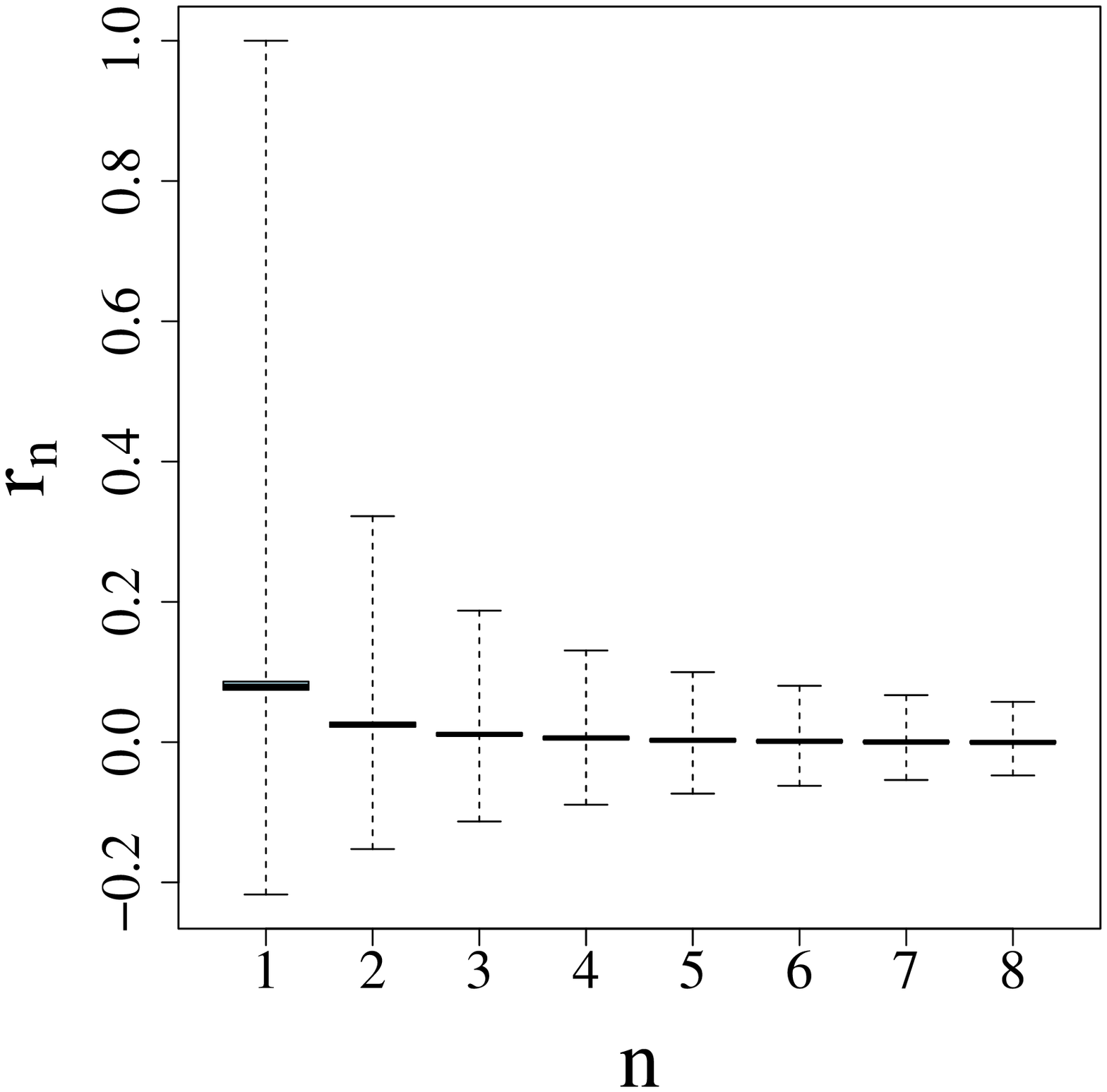}} \hfill
   \resizebox{0.44\hsize}{!}{\includegraphics[keepaspectratio, angle=-90]{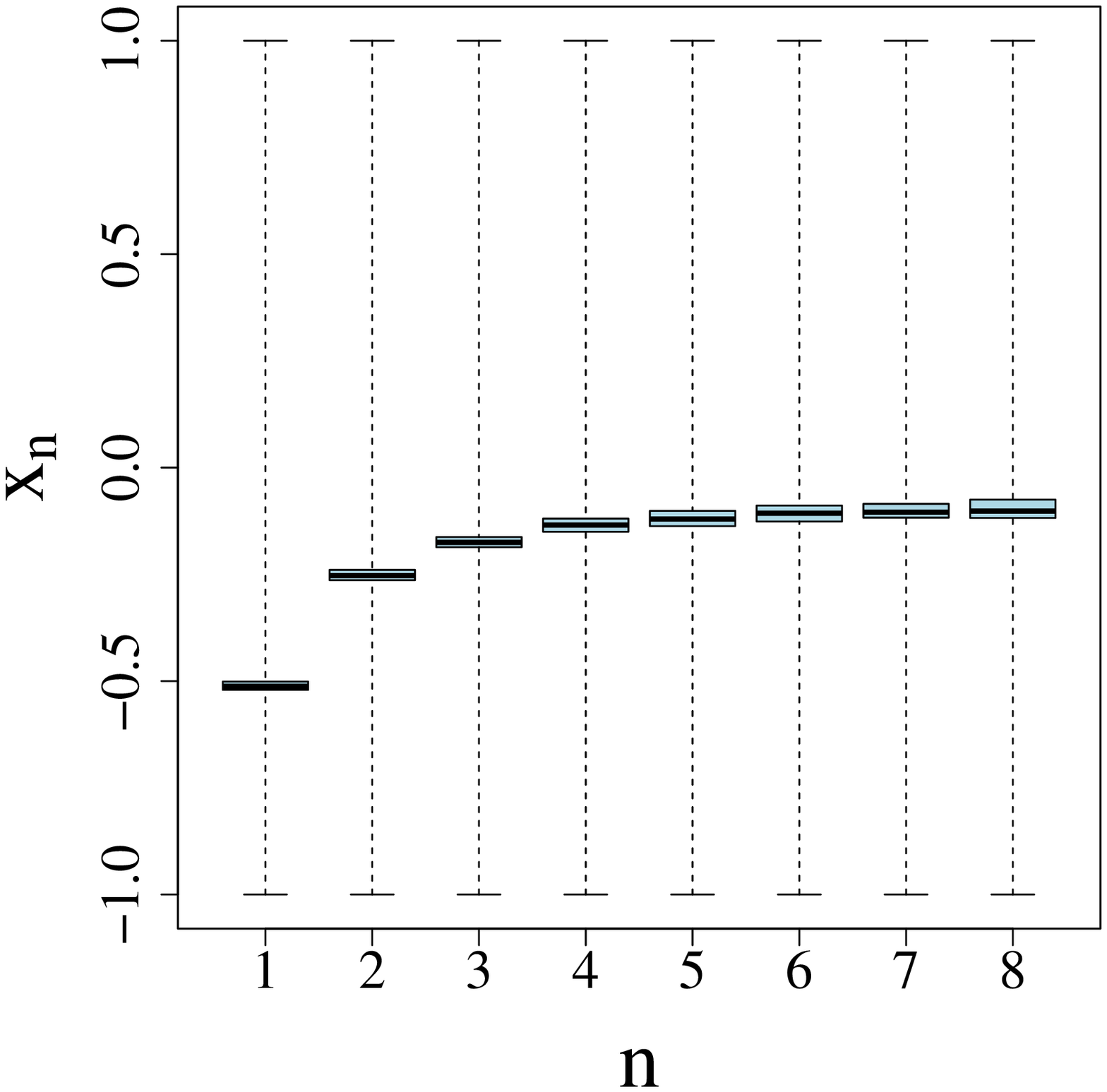}}
   \caption{Box-and-whisker plots for the correlation coefficients $r_n$ and the transformed values $x_n$, as defined in \protect\autoref{eq:r_to_x}, where each data point shows the upper and lower quartile (edges of the box), median (line inside the box), and mean upper and lower boundaries (whiskers) for the 125 realizations measured from the simulation subcubes. These use the same lags, random catalog size, and estimator as before (see text and previous figure captions for details).}
   \label{fig:boxplot_r_x}
\end{figure*}

\autoref{fig:boxplot_r_x} shows box-and-whisker plots of $r_n$ and $x_n$ at all eight lags $n$, where we use the same lags and random catalog size as before, as well as the LS estimator.
\begin{figure*}
   \resizebox{0.44\hsize}{!}{\includegraphics[keepaspectratio, angle=-90]{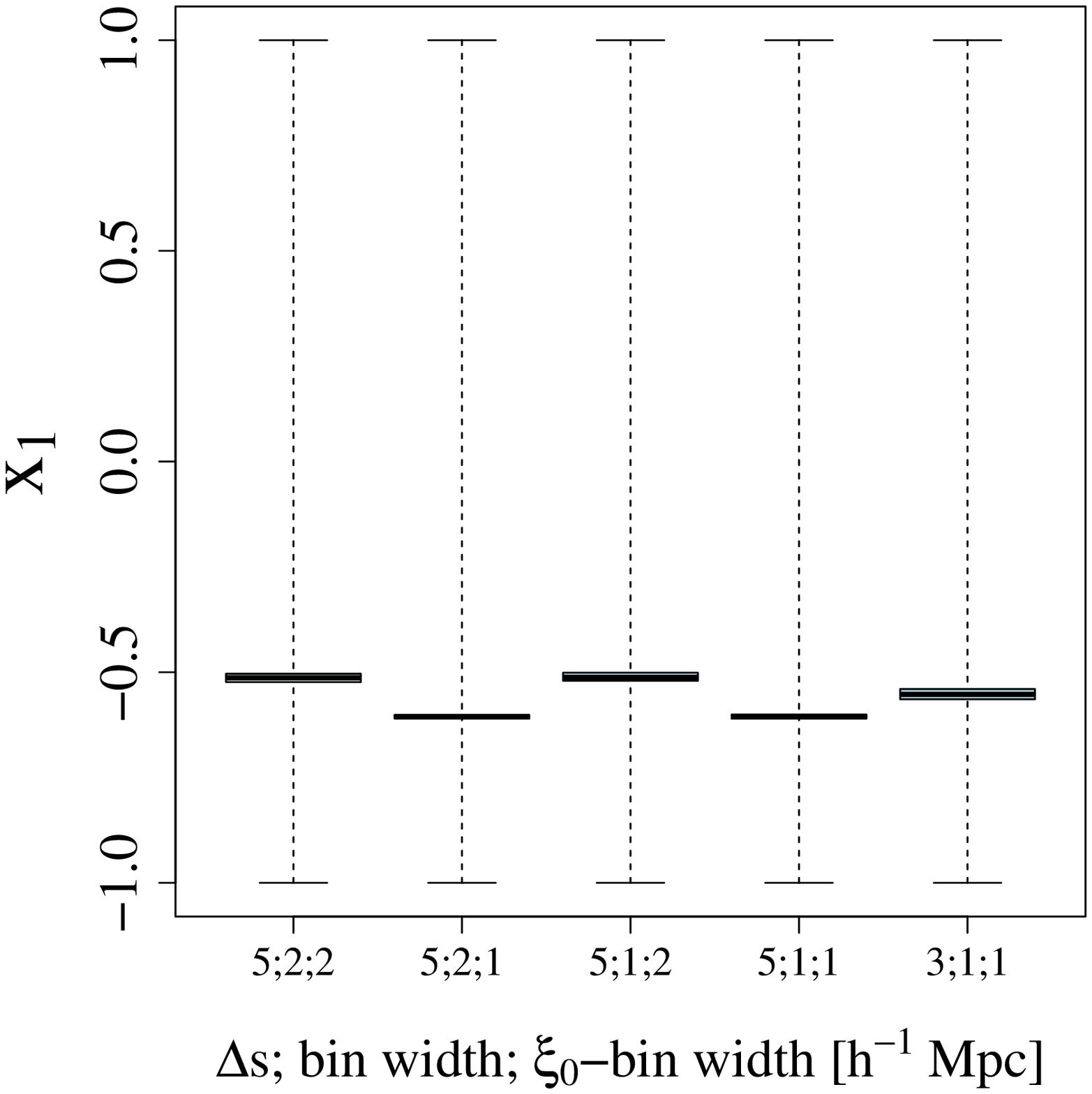}} \hfill
   \resizebox{0.44\hsize}{!}{\includegraphics[keepaspectratio, angle=-90]{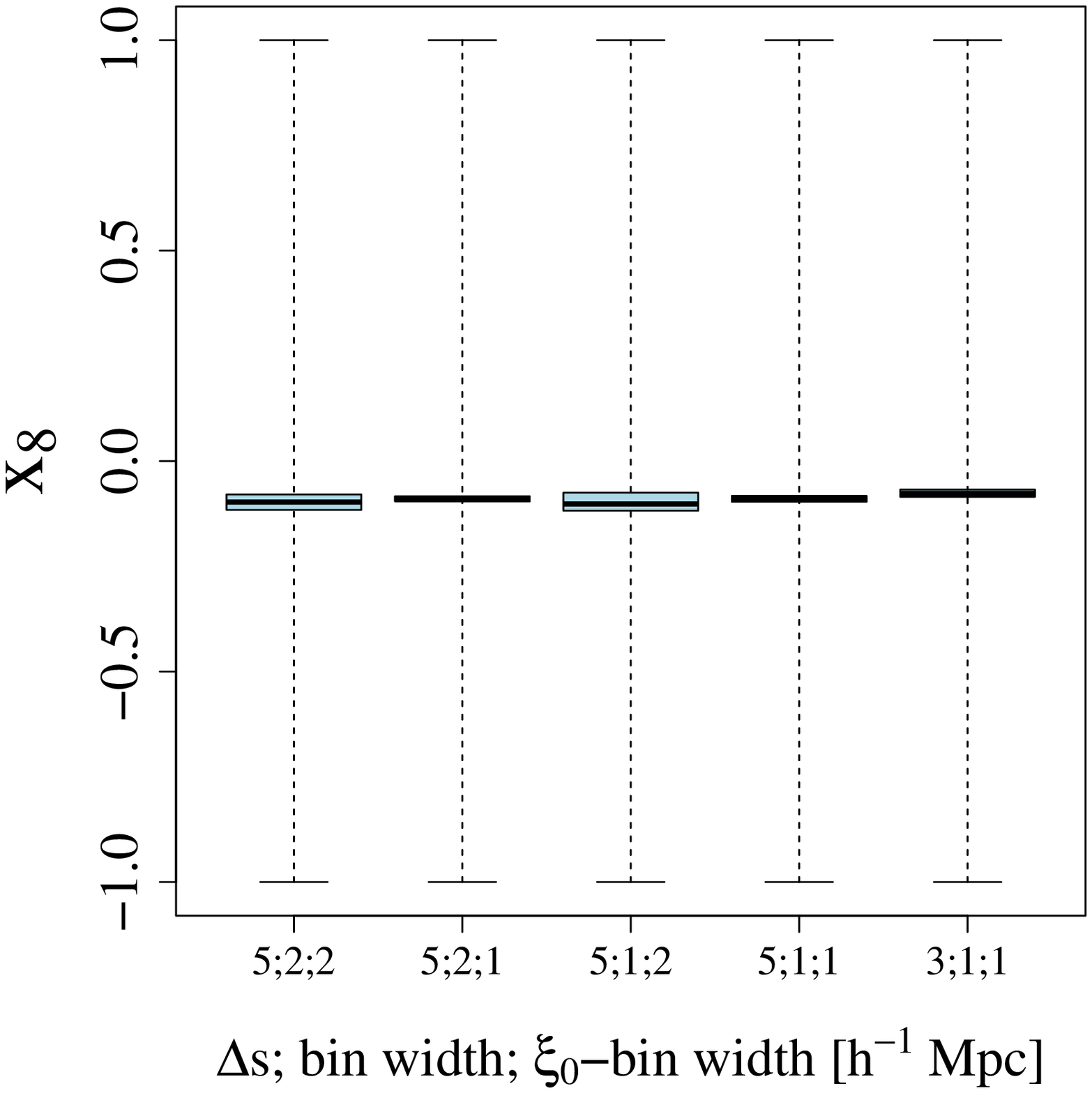}}
   \caption{Box-and-whisker plots of the transformed correlation coefficients at smallest and largest lag for varying lag separation and bin width. The triple labeling of each distribution gives the lag separation $\Delta s$, the bin width of pair separations at non-zero lag (i.e., for $\xi_1 \ldots\xi_8$), and the width of the $\xi_0$-bin. For example, for the second case shown in each panel, we measure $\xi_0$ from halo pairs with separations from 0 to $\unit[1\ \hinv]{Mpc}$, $\xi_1$ from those with separations from 4 to $\unit[6\ \hinv]{Mpc}$, $\xi_2$ from 9 to $\unit[11\ \hinv]{Mpc}$, and so on.}
   \label{fig:boxplot_x_diff_lags}
\end{figure*}
It can be seen that the constraints are clearly obeyed, and although the distributions becoming broader for increasing lags, the boxes showing the upper and lower quartiles only occupy a small portion of the allowed region. The distributions are not necessarily centered within the allowed region, which is not surprising, since their exact shape and position also depend on the underlying power spectrum.
The choice of estimator has barely any impact on the widths and positions of the distributions within the allowed region, which is to be expected, since \autoref{fig:xi_of_s_est} already illustrates that the different estimators yield quite similar results.

As mentioned at the beginning of this section, the main influence on the variance of the distributions in $\xi$, and correspondingly in $r$- and $x$-space, seems to be the width of the $\xi_0$-bin. In \autoref{fig:boxplot_x_diff_lags} we investigate this observation and also study the impact of the choice of the separation between the lags at which we measure $\xi$.
In the two panels of the figure, we show box-and-whisker plots of the transformed correlation coefficients at smallest and largest lag, i.e., $x_1$ and $x_8$, and we vary the separation $\Delta s$ of the lags as well as the bin widths of the pair separations used to measure $\xi_0$ and the correlation functions at non-zero lag, $\xi_1\ldots\xi_8$.
In the case of the four left-most distributions in each panel, we use a lag separation of $\Delta s=\unit[5\ \hinv]{Mpc}$, where we adapt a bin width of $\unit[1\ \hinv]{Mpc}$ for $\xi_1\ldots\xi_8$ for the first and second distribution, and a bin width of $\unit[2\ \hinv]{Mpc}$ for the third and fourth ones. In both cases, we separately use a narrow and a broad bin width for the measurement of $\xi_0$ (also 1 and $\unit[2\ \hinv]{Mpc}$).
The figure illustrates that the width of the distributions of $x_n$ is mainly determined by the $\xi_0$-bin size, whereas the width of the bins for $\xi_n$ at lags $n>0$ barely has any influence.
This is due to the structure of the quasi-Gaussian transformation, where $\xi_0$ appears in every correlation coefficient $r_n$, and, as a result, in the computation of every lower and upper bound.
The impact of the width of the $\xi_0$-bin is particularly strong for small-lag distributions, which it also shifts, as can be seen from the distribution of $x_1$. In particular, this shift is larger than  a case where we measure $\xi_n$ at different lags altogether, as illustrated for a lag separation of $\Delta s=\unit[3\ \hinv]{Mpc}$ in the fifth distribution shown in the figure.
In this context, it is important to stress that the problem of how to measure $\xi_0$ in practice is well-known since, in most applications, it is difficult to measure $\xi$ at very small lags. As we  show, however, this poses a particularly difficult challenge when analyzing measured correlation functions in a quasi-Gaussian framework, since here, the exact determination of $\xi_0$ is vital; the auto-correlation function enters everywhere, because one would always transform $\xi$ to $y$ (or at least to $r$) for an analysis involving the constraints.
\begin{figure*}
   \resizebox{0.44\hsize}{!}{\includegraphics[keepaspectratio, angle=-90]{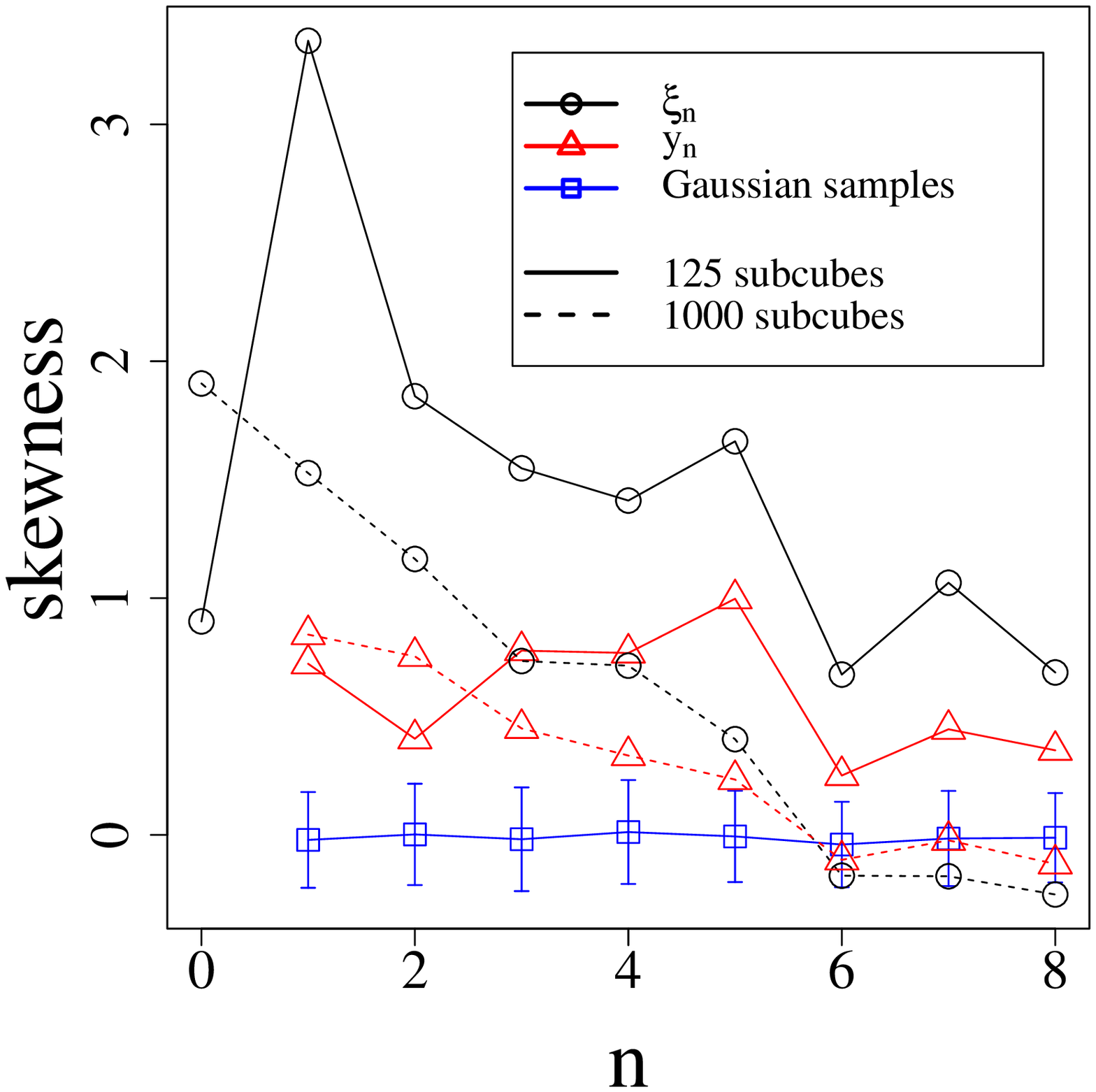}}\hfill
   \resizebox{0.44\hsize}{!}{\includegraphics[keepaspectratio, angle=-90]{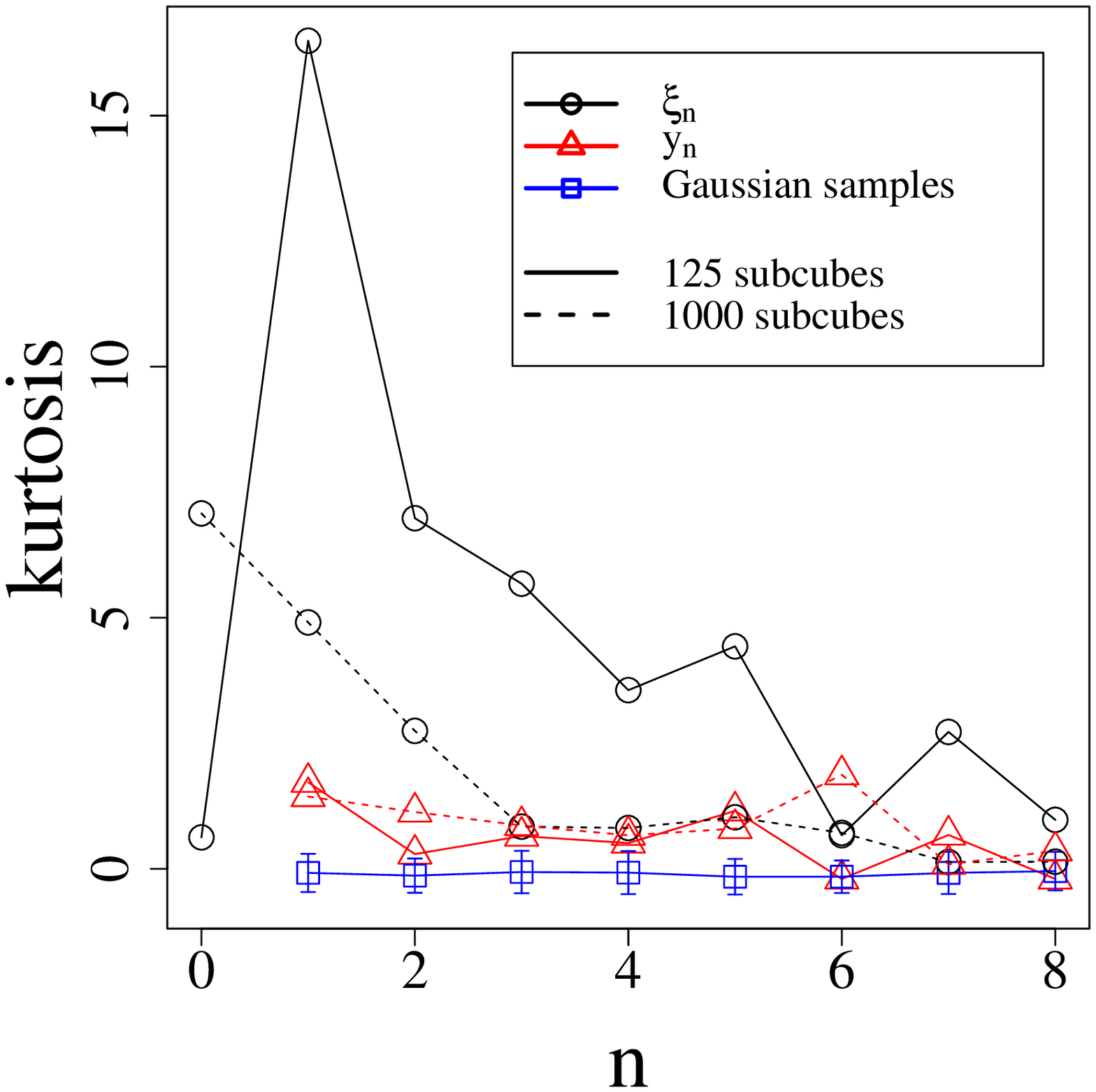}}
   \caption{Test for the univariate Gaussianity of the $\lbrace\xi\rbrace$-and $\lbrace y\rbrace$-samples obtained from the Millennium Simulation, using a lag separation of $\Delta s=\unit[5\ \hinv]{Mpc}$ and bin widths of $\unit[1\ \hinv]{Mpc}$ for all $\xi_n$, including $\xi_0$. The black circles and red triangles show the univariate skewness and kurtosis of the distributions $p(\xi_n)$ and $p(y_n)$, computed over 125 (solid curves) and 1000 (dashed curves) subcubes of the simulation volume. For the blue curves, we draw 100 Gaussian samples with the same mean, covariance matrix and sample size as the distributions $p(y_n)$ in the case of 125 subcubes, and plot the mean and standard deviation of their skewness and kurtosis.}
   \label{fig:skew_kurt_uni}
\end{figure*}

In summary, all correlation functions we measured from the Millennium Simulation are quite far away from the edge of the allowed region. This finding seems to hold irrespective of the choice of estimator, lags, etc., providing that $\xi$ is measured in a `sensible' way. As an example, using very small random catalog sizes does indeed yield single realizations outside the allowed region.

\subsection{Quality of the Gaussian approximation in \texorpdfstring{$\xi$}{xi} and \texorpdfstring{$y$}{y}-space}
In this section, we use the correlation function samples measured from the Millennium Simulation to assess the quality of a quasi-Gaussian approach. 
Similar to the tests shown in \WS{} for simulated correlation functions, we transform $\xi$ to $y$ as defined in \autoref{eq:r_to_y} and test the Gaussianity of the distributions in $y$ and $\xi$, because the Gaussianity in $y$-space is a central prerequisite for the accuracy of the quasi-Gaussian likelihood.

While it would be preferable to  assess the quality of the quasi-Gaussian approximation directly, i.e., to check how well the quasi-Gaussian PDF agrees with $p(\xi),$ as obtained from the Millennium Simulation, computing the quasi-Gaussian PDF still requires measuring the underlying power spectrum, which is beyond the scope of this work. In real life, however, one would usually transform the measured correlation function to $y$-space to perform a Bayesian analysis and, thus, the Gaussianity of $p(y)$ is pivotal. Even so, knowledge about the underlying power spectrum would still be required  to make use of the analytically known $p(\xi_0)$.
\begin{figure*}
   \resizebox{0.44\hsize}{!}{\includegraphics[keepaspectratio, angle=-90]{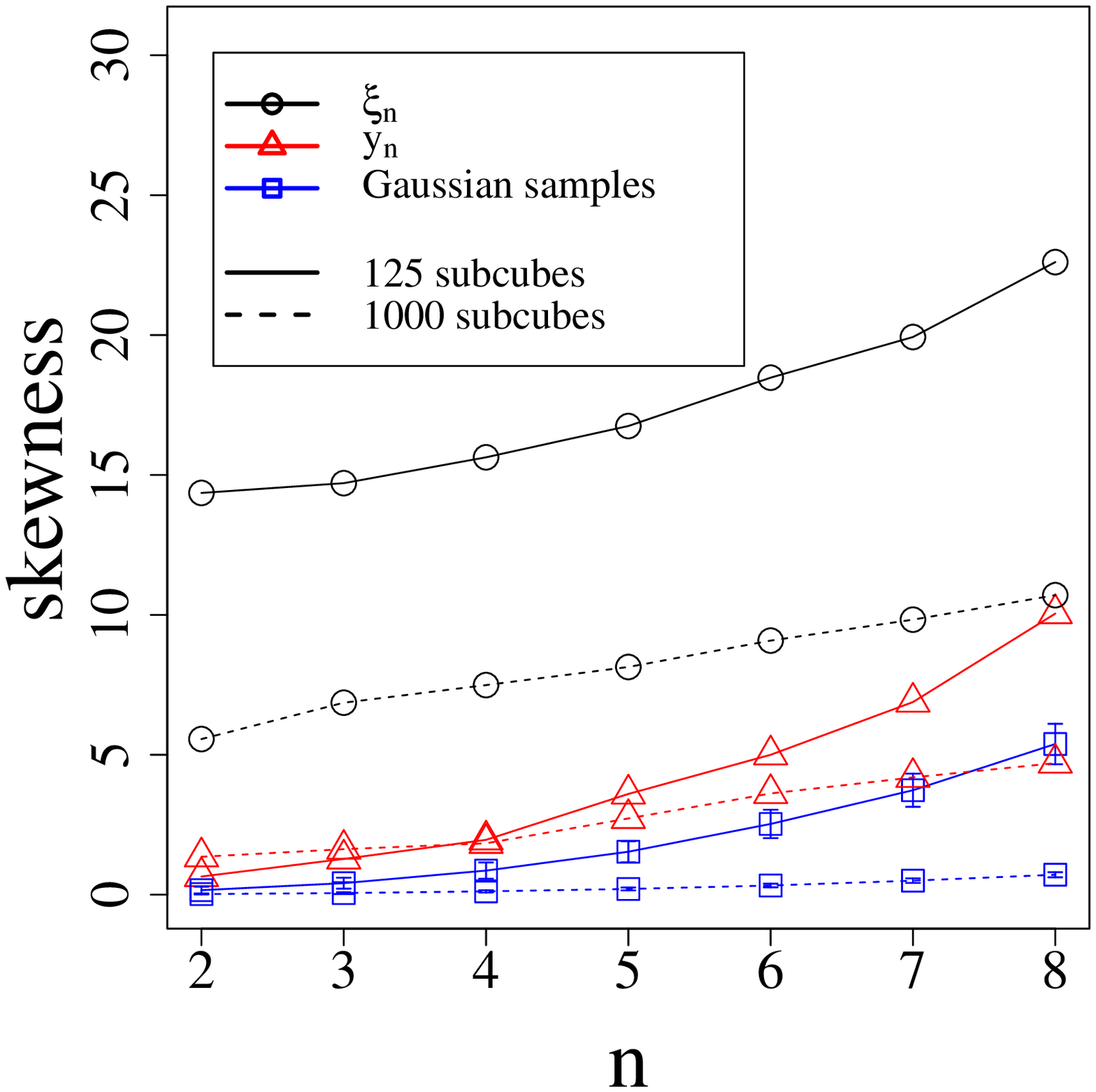}}\hfill
   \resizebox{0.44\hsize}{!}{\includegraphics[keepaspectratio, angle=-90]{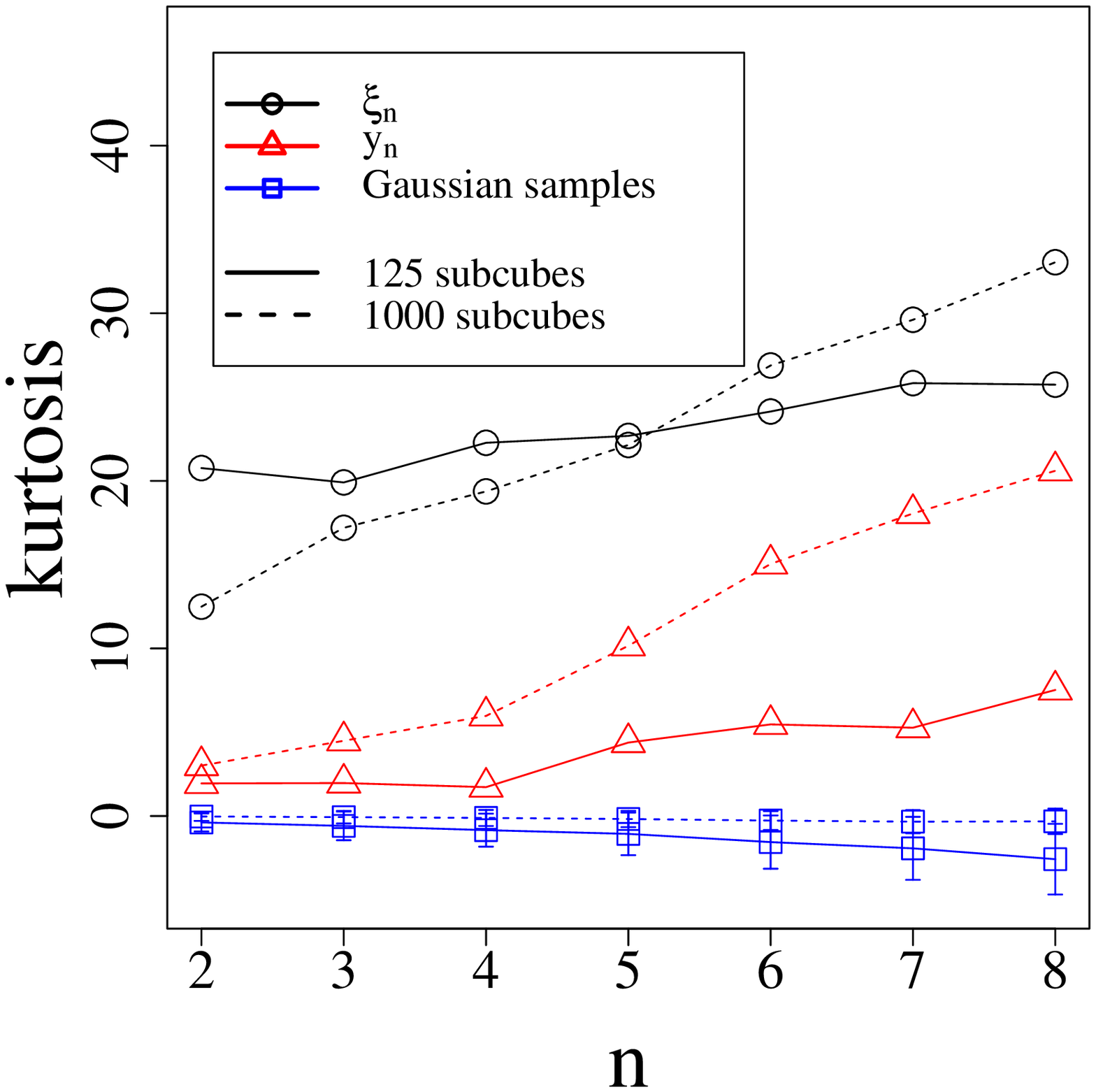}}
   \caption{Multivariate Mardia's skewness and kurtosis of the $n$-variate distributions of the $\lbrace\xi\rbrace$-and $\lbrace y\rbrace$-samples, obtained from the Millennium Simulation, and of corresponding Gaussian samples, using the same parameters as before (see previous figure caption). In contrast to the previous figure, we plot two curves for Gaussian samples: For the solid (dashed) curve, we draw Gaussian samples with the same mean, covariance matrix, and sample size as the corresponding distributions in $y$-space for the case of 125 (1000) subcubes; the blue squares and error bars show the mean and standard deviation of the skewness and kurtosis of the 100 samples.}
   \label{fig:skew_kurt_multi}
\end{figure*}

In the literature, various tests for Gaussianity exist. In this study, we focus on the calculation of moments; in particular, we compute the skewness and kurtosis, which are defined in such a way that they are zero for a Gaussian distribution. 
In the univariate case, the skewness $\gamma$ of a distribution $p(x)$ reads
\begin{equation}
 \gamma=\left\langle\frac{(x-\mu)^3}{\sigma^3}\right\rangle\equiv\frac{m_3}{m_2^{3/2}},
 \label{eq:skewness}
\end{equation}
where $m_i=\langle (x-\mu)^i\rangle$ denotes the central $i$th-order moment. Thus, $\gamma$ is essentially the (renormalized) third-order moment, and the kurtosis
\begin{equation}
 \kappa=\left\langle\frac{(x-\mu)^4}{\sigma^4}\right\rangle-3\equiv\frac{m_4}{m_2^2}-3
 \label{eq:kurtosis}
\end{equation}
is closely related to the fourth-order moment.
In the multivariate case, we use the definitions established by \cite{bib_mardia_1970, bib_mardia_1974}, who define the skewness of a $d$-variate distribution as
\begin{equation}
 \gamma_d=\frac{1}{n^2}\sum_{i=1}^n\sum_{j=1}^n \left\lbrace \transpose{\left(\vec x_i -\vec\mu\right)} C^{-1} \left(\vec x_j -\vec\mu\right)\right\rbrace^3,
\end{equation}
where $n$ is the sample size, and $\vec\mu$ and $C$ are the sample mean and covariance matrix. The kurtosis measure reads
\begin{equation}
 \kappa_d=\frac{1}{n}\sum_{i=1}^n \left\lbrace \transpose{\left(\vec x_i -\vec\mu\right)} C^{-1} \left(\vec x_i -\vec\mu\right)\right\rbrace^2 -d(d+2),
\end{equation}
where we subtract the last term to ensure that a perfectly Gaussian sample yields $\kappa_d=0$.

To test the impact of the quasi-Gaussian transformation on Gaussianity, we transform each realization of the correlation function (measured for eight lags of separation $\Delta s=\unit[5\ \hinv]{Mpc}$ with bins of width $\unit[1\ \hinv]{Mpc}$ for $\xi_0\ldots\xi_8$) to $y$ and compute skewness and kurtosis of the distributions in $\xi$- and $y$-space.
Analogously to our tests in \WS{}, we also draw Gaussian samples with the same mean and covariance matrix as our samples $\lbrace y\rbrace$, both for comparison and to account for small sample sizes.
The results for the univariate distributions are plotted in \autoref{fig:skew_kurt_uni}. Here, we show the skewness and kurtosis of the distributions $p(\xi_0),\ldots, p(\xi_8)$ and $p(y_1),\ldots, p(y_8)$; for the solid lines, we sliced the simulation volume into 125 subcubes, whereas the dashed lines correspond to the case of 1000 subcubes.
By way of  comparison, the blue curves show the skewness and kurtosis of corresponding Gaussian samples, where for the sake of clarity, we only plot the curves for 125 subcubes. Because skewness
and kurtosis fluctuate quite significantly for this small sample size, we draw 100 Gaussian samples of size 125 and compute the skewness and kurtosis of each sample;  the blue squares and error bars show the mean and standard deviation of the skewness and kurtosis of the 100 samples.
We only include the purely Gaussian samples for comparison and to check how close to zero the measured skewness and kurtosis are for these small sample sizes. Specifically, they are not meant to provide any insight into the absolute scale of the non-Gaussianity observed in $\xi$.
Evidently, the distributions in $y$ are far more Gaussian than those in $\xi$ (with the exception of $p(\xi_0)$ in the case of 125 subcubes) and, of particular note, show a kurtosis comparable to the Gaussian samples.

Since the Gaussianity of the univariate distributions does not imply Gaussianity of the multivariate PDFs, we also compute the moments of the $n$-variate distributions $p(\xi_0,\ldots,\xi_{n-1})$, $p(y_1,\ldots,y_n)$ and of corresponding multivariate Gaussian samples, and plot them as functions of $n$ , as  in \autoref{fig:skew_kurt_multi}.
Here, we show the results for corresponding Gaussian samples for both 125 and 1000 subcubes, where the plotted values and error bars  are the mean and standard deviation of the skewness and kurtosis computed over 100 Gaussian samples. While the multivariate moments of the Gaussian samples of size 125 are not consistent with zero, this is indeed the case for the larger sample size of 1000.
For the dashed curves, i.e., the case of 1000 subcubes, in the case of higher $n$, the integral constraint has a non-negligible impact on the measured correlation functions, as explained in the previous section. As a consequence, the corresponding skewness and kurtosis results should only be considered quantitatively.
Nevertheless, it is clear that the difference between the level of Gaussianity in $\xi$- and $y$-space becomes even larger for the multivariate case, reaching about one order of magnitude in $\gamma$ and $\kappa$ in the case of 125 subcubes.

\begin{figure*}
   \resizebox{0.44\hsize}{!}{\includegraphics[keepaspectratio, angle=-90]{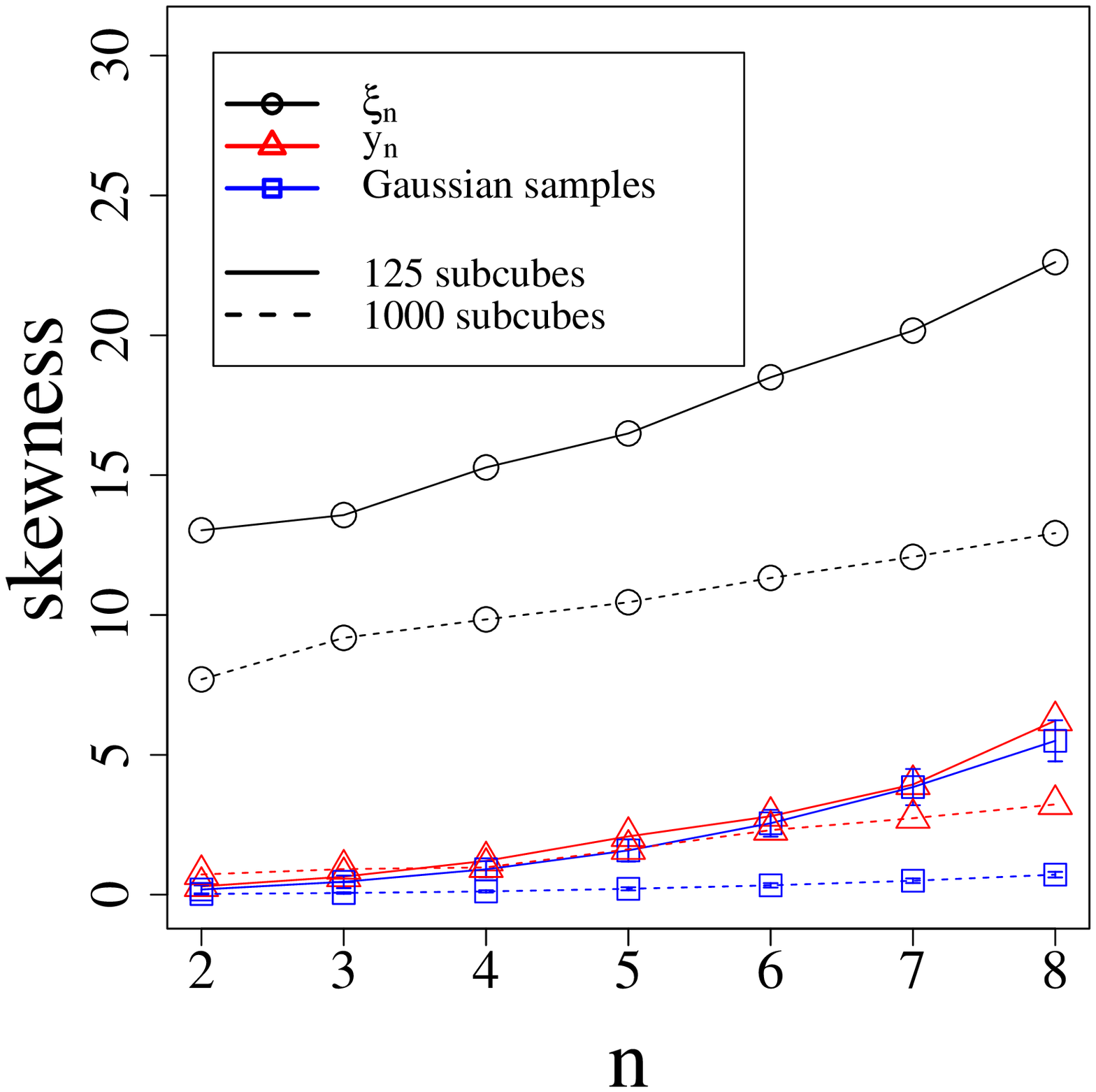}}\hfill
   \resizebox{0.44\hsize}{!}{\includegraphics[keepaspectratio, angle=-90]{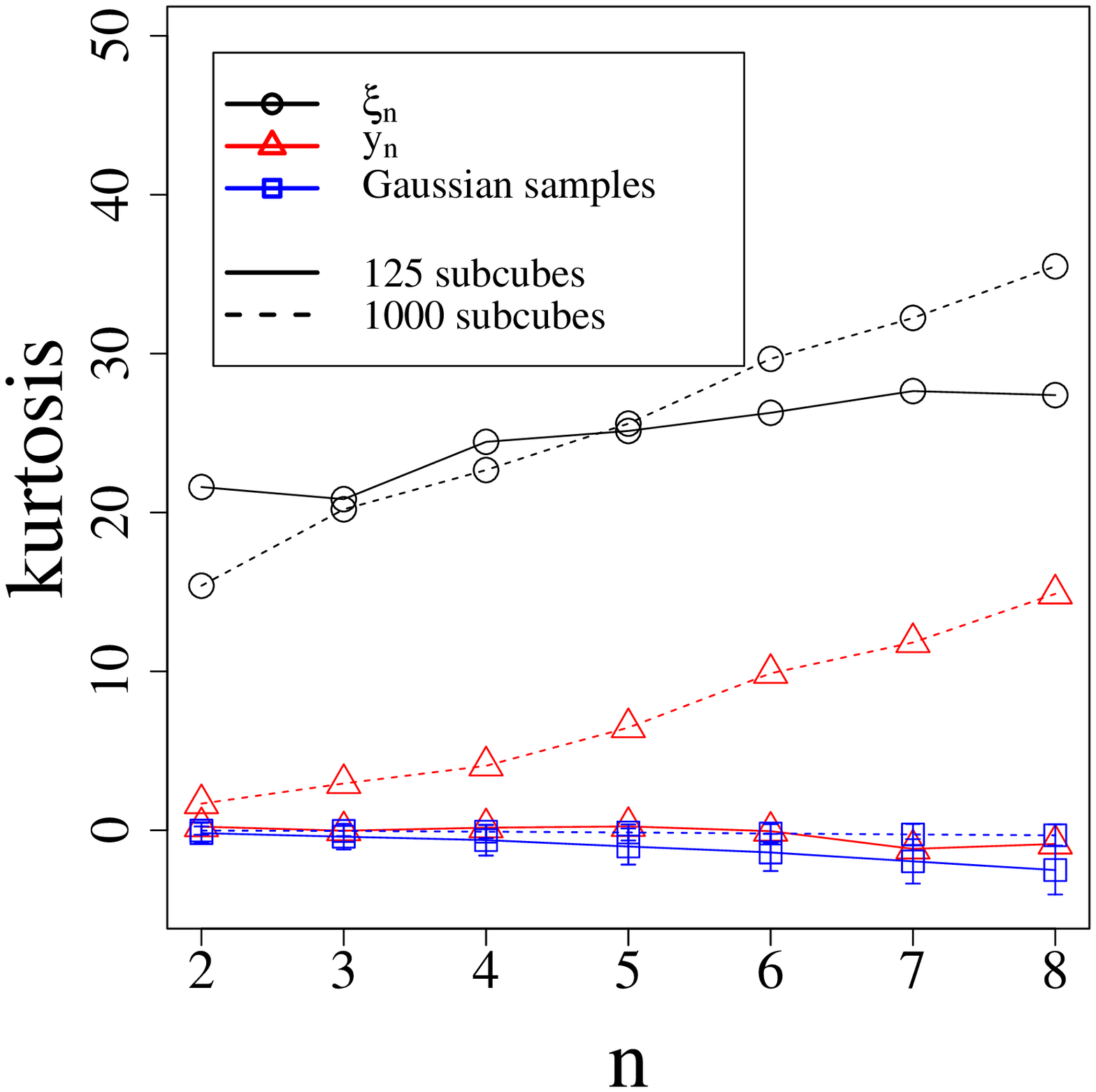}}
   \caption{Multivariate skewness and kurtosis of the $\lbrace\xi\rbrace$-and $\lbrace y\rbrace$-samples obtained from the Millennium Simulation and of corresponding Gaussian samples. When compared to the previous figure, we adapt a broader $\xi_0$-bin,  i.e., we measure the auto-correlation function from all halo pairs with pair separations from 0 to $\unit[2\ \hinv]{Mpc}$.}
   \label{fig:skew_kurt_multi_broad_xi0}
\end{figure*}

As we demonstrated in the previous section, the width of the $\xi_0$-bin, i.e., the range of pair separations used to measure the auto-correlation function, has an impact on the distributions of the correlation coefficients, and thus on those of the $y_n$. Accordingly, we vary the $\xi_0$-bin width and again study the multivariate moments of the corresponding distributions. \autoref{fig:skew_kurt_multi_broad_xi0} shows a similar plot to \autoref{fig:skew_kurt_multi}, but we use a $\xi_0$-bin width of 2 instead of $\unit[1\ \hinv]{Mpc}$.
As it turns out, this yields distributions in $y$-space which are almost perfectly Gaussian, at least in the case of 125 subcubes, where their moments are hardly distinguishable from those of the corresponding Gaussian samples with same sample size.
As before, it seems that the width of the $\xi_0$-bin has a far higher impact on the results than the bin widths for $\xi_1\ldots\xi_8. $ Actually, using bins of $\unit[2\ \hinv]{Mpc}$ for the higher-lag correlation functions barely influences the outcome.

In summary, all tests shown in this section indicate that the distributions in $y$-space are far more Gaussian than those in $\xi$, and in some cases even have skewness and kurtosis comparable to those of Gaussian samples of the same size. This demonstrates the validity of the quasi-Gaussian approach independent of the specific parameters used to measure the correlation function.


\section{Conclusions and outlook}
\label{sec:conclusions}
Building on \SH{}, we have developed numerical methods to compute the fundamental constraints on correlation functions. We  have shown  these methods, which are applicable also in the case of two- and three-dimensional random fields, to be robust and precise, since the numerical computation of the constraints for the one-dimensional case reproduces the analytically known bounds.
We then   applied our results to samples of correlation functions measured from the halo catalog of the Millennium Simulation. After discussing some challenges in the measurement of $\xi$, such as the choice of random catalog size and lag separation, as well as the question of how to overcome the integral constraint, we  have shown that the correlation functions measured from the simulation very clearly obey the constraints.
Even though all measured correlation functions lie far away from the edges of the allowed region, we  have demonstrated that the quasi-Gaussian quantity $y$ yields significantly smaller non-Gaussian signatures (i.e., skewness and kurtosis) than the original correlation function $\xi$, giving further support to the claim that the quasi-Gaussian approximation for the correlation function likelihood, introduced in \WS,{} is a far better description than the Gaussian one.

As a brief outlook on possible future work, one vital improvements would be to bypass the current limitation to eight lags in the numerical computation of the constraints, since modern astronomical observations usually measure $\xi$ at far more lags. 
Furthermore, the performance of the quasi-Gaussian likelihood in the three-dimensional case should be assessed and compared to the classical Gaussian approach. While this would be testable on the samples of correlation functions measured from the Millennium Simulation, the most significant advance would be the application of our methods to real data, and an investigation of their impact on cosmological parameter estimation.
Aside from the current limitation to only eight lags, this would pose additional challenges, depending on the area of application: In the case of a redshift survey, for example, different constraints on the correlation function, measured along and perpendicular to the line-of-sight, would hold as  a result of redshift space distortions. Nonetheless, the constraints on correlation functions of three-dimensional random fields are, in principle, treatable, despite open challenges and room for improvements. Taking these things into consideration, this work opens up a vast field of applications where Gaussian likelihoods for $\xi$ have previously been used.


\begin{acknowledgements}

We would like to thank Cristiano Porciani for useful discussions and help.
We also thank our anonymous referee for constructive comments and suggestions.
The Millennium Simulation databases used in this paper, and the web application providing online access to them, were constructed as part of the activities of the German Astrophysical Virtual Observatory (GAVO).
This work was in part supported by the Deutsche Forschungsgemeinschaft, under the project SCHN~342/11.

\end{acknowledgements}


\bibliographystyle{aa}   
\bibliography{bib_ccfmill} 


\end{document}